\documentclass{emulateapj}
\usepackage{graphicx}
\begin{document}
\title{Wide Field Near-Infrared Photometry of 12 Galactic Globular Clusters: Observations Versus Models on the Red Giant Branch}
\author{
Roger E. Cohen\altaffilmark{1},
Maren Hempel\altaffilmark{2},
Francesco Mauro\altaffilmark{3},
Douglas Geisler\altaffilmark{1},
Javier Alonso-Garcia\altaffilmark{4,3},
Karen Kinemuchi\altaffilmark{5}
}

\altaffiltext{1}{Departamento de Astronom\'{i}a, Universidad de Concepci\'{o}n, Casilla 160-C, Concepci\'{o}n, Chile}
\altaffiltext{2}{Instituto de Astrof\'isica, Facultad de F\'isica, Pontificia
  Universidad Cat\'olica de Chile, Av.~Vicu\~na Mackenna 4860, 782-0436 Macul,
  Santiago, Chile}
\altaffiltext{3}{Instituto Milenio de Astrof\'isica, Santiago, Chile}
\altaffiltext{4}{Unidad de Astronom\'ia, Universidad de Antofagasta,
  Avda. U. de Antofagasta 02800, Antofagasta, Chile}
\altaffiltext{5}{Apache Point Observatory/NMSU, Sunspot, NM 88349, USA}
\date{\today}

\begin{abstract}
We present wide field near-infrared photometry of 12 Galactic globular
clusters, typically extending from the tip of the cluster 
red giant branch (RGB) to
the main sequence turnoff.  
Using recent homogenous values of cluster distance, reddening and
metallicity, the resulting photometry is directly compared 
to the predictions of several recent libraries of stellar evolutionary models. 
Of the sets of models investigated, Dartmouth and Victoria-Regina models 
best reproduce the observed RGB morphology, albeit with offsets in 
$J-K_{S}$ color
which vary in their significance in light of all sources of
observational uncertainty.  
Therefore, we also present newly recalibrated relations between
near-IR photometric indices describing the upper RGB  
versus cluster iron abundance as well as global metallicity.
The influence of enhancements in alpha
elements and helium are analyzed, finding that the former affect the
morphology of the upper RGB in accord with model predictions.  
Meanwhile, the empirical relations we
derive are in good agreement with previous results, and minor discrepancies
can likely be attributed to differences in the assumed cluster distances and
reddenings.  In addition, we present measurements of the horizontal branch (HB) and
RGB bump magnitudes, finding a non-negligible dependence of the
near-IR HB magnitude on cluster metallicity.  Lastly, we discuss the influence 
of assumed cluster distances, reddenings and metallicities on our results, 
finding that our empirical relations are generally insensitive to these 
factors to within their uncertainties.  

\end{abstract}

\keywords{globular clusters: general --- globular clusters: individual(NGC
  104, NGC 288, NGC 362, NGC 1261, NGC 1851, NGC 2808, NGC 4833, NGC 5927, 
NGC 6304, NGC 6496, NGC 6584, NGC 7099) --- stars: infrared}

\section{Introduction}

Galactic globular clusters (GGCs) play a critical role as templates of evolved 
stellar populations, functioning as testbeds for stellar evolutionary models.  
At optical wavelengths, huge advances in our understanding of GGCs have been
made via extensive photometric surveys of large samples of GGCs both from
space \citep{piottosnapshot,sarajedini07,piotto14}, and from the ground 
\citep[e.g.][]{rosenberg,stetson2000}.  The homogeneity of these databases has
facilitated a variety of comparisons between precise photometry and existing
evolutionary models (e.g.~\citealt{marinfranch}; Dotter et al.~2010, hereafter \citealt{d10}; VandenBerg et
al.~2013, hereafter \citealt{v13}).  
However, at near-infrared (near-IR) wavelengths, such comparisons
remain few despite a growing database of observations.  
\citet{brasseur} compared $VJK_{S}$ photometry of 6 GGCs (and the
old open cluster NGC 6791) to the predictions of the latest Victoria-Regina
models \citep{v14}, finding that the models fail to
reproduce the observed morphology of cluster RGBs, at
least at lower metallicities.  
On the other hand, the Victoria-Regina models seem to function well
in the near-IR when compared to M4 and NGC 6723 \citep{hendricks} and the main
sequence of NGC 3201 \citep{bonoknee}.  Other direct comparisons between
near-IR GGC isochrones and photometry were performed by \citet{v04obs,v04abs}
using then-recent models \citep{caloi,salaris97,scl97} and bolometric
corrections \citep{montegriffo}, and \citet{salaris47tuc} compared
the observed HB and RGB bump and luminosity function of 47 Tuc 
to $\alpha$-enhanced
BaSTI models \citep{basti}.  The lack of existing comparisons between
isochrones and data in the near-IR is not due to a lack of model predictions,
and in fact \citet{sg02} tabulate the predicted mean absolute $K$
magnitude $M_{K}$ of cluster HBs as a function of age
and metallicity.  However, at ages typical of GGCs ($>$10 Gyr) these
predictions have never been tested thoroughly as the combination of
sufficiently accurate ages and $K$-band photometry was lacking.

The systematic observation of GGCs in the near-IR has been undertaken largely
by E. Valenti and collaborators, yielding a database of near-IR photometry of
optically well-studied GGCs, including relations between observed photometric
features and metallicity (\citealt{ferraro00,v04obs,v04abs,ferraro06}) in addition to
\citet{chobump} and \citet{jura2mass}, who undertook similar analyses using photometry from the Two Micron All Sky Survey (2MASS; \citealt{skrutskie}). 
These studies have since served as a \textit{de facto} template for old ($>$10 Gyr) stellar populations in the near-IR, and have therefore seen numerous applications.  These include the measurement of photometric distances, reddenings and metallicities for many poorly studied GGCs located towards the Galactic bulge \citep{kim,chun,v10}, Local Group dwarf galaxies \citep[e.g.][]{gorskitrgb,heldleo}, and reddening and metallicity maps of the Galactic bulge \citep[e.g.][]{gonzalezmaps,gonzalezfeh}.

Meanwhile, 
the proliferation of space-based photometric GGC surveys has yielded 
homogenous databases of GGC cluster distances, reddenings 
and ages \citep[e.g.][]{d10,v13} while 
high resolution multi-object spectroscopic campaigns have produced
detailed chemical abundance studies which include large homogenous GGC abundance databases of $[Fe/H]$ (Carretta et al.~2009a, hereafter \citealt{c09}) as well as light and $\alpha$-elements \citep[e.g][]{c07,c09b,c09c,c10,c10b}. 
With the goal of leveraging these improvements together, we present
near-IR photometry of a sample of optically well-studied GGCs, which we employ
to reexamine relations between photometric features and metallicity.  We then
compare our observations to the predictions of several sets of evolutionary
models in near-infrared colors for the first time.

In the next section, we describe the observations and data reduction.  In section three, we present color-magnitude diagrams and empirical fiducial sequences for all of our target clusters.  
Section four contains a comparison between the observed RGB fiducial sequences and predictions of
five sets of evolutionary models, as well as updated calibrations of
photometric indices versus cluster metallicity. 
In section five, we present observed magnitudes of the horizontal branch and
red giant branch bump, as well as empirical relations between the near-IR
absolute magnitude of these features versus cluster metallicity.
Section six consists of a discussion of current uncertainties in 
cluster distance, reddening and metallicity, and their influence on our empirical calibrations, and in the final section we summarize our results.

\section{Observational Data}
\subsection{Observations and Pre-Processing}

Observations of our 12 target clusters
were obtained with the 
Infrared Side Port Imager (ISPI) mounted
on the 4m Blanco telescope at Cerro Tololo Inter-American Observatory.  The
HAWAII-2 2048$\times$2048 pix detector has 0.305$\arcsec$/pixel, giving a field of view 10.25 arcmin per side.  Imaging was obtained in the $J$ and $K_{S}$
filters 
over the course of three runs between 2008 and 2010, with median seeing
ranging between 0.8\arcsec and 1.4\arcsec, and a log of the
observations is given in Table \ref{logtab}.  In order to mitigate the
effects of persistence, saturation, and cosmetic defects, a two-step (for the
2008 run) or five-step (all other runs) dither pattern was employed for each
cluster, where each individual 60s exposure in the dither pattern is comprised
of 3-12 coadds (as given
in the last two columns of Table \ref{logtab}) to optimize dynamic range.  As
near-infrared imaging is typically limited by the sky background, which can
vary both spatially (on scales smaller than the detector field of view) and
temporally (on timescales of minutes), careful subtraction of this sky
background is critical to mitigating photometric errors.  Therefore, each
on-target dither sequence for each cluster was intercalated with identical
sequences targeted at an offset sky field located about $\sim$15$\arcmin$
away.  These offset sequences were used to create a combined sky frame
corresponding to each on-target sequence.  
Preprocessing, including the construction of bad pixel masks, bias
subtraction, and flat fielding, was accomplished using IRAF\footnote{IRAF is distributed by the National Optical Astronomy Observatories, which are operated by the Association of Universities for Research in Astronomy, Inc., under cooperative agreement with the National Science Foundation.} tasks customized for ISPI within the \texttt{cirred} package\footnote{See \url{http://www.ctio.noao.edu/noao/content/ISPI-Data-Reduction}}.  Next, 
sky subtraction was optimized by using a multi-step iterative procedure: 
Sources are rejected to fit an initial sky
background, before median combining and subtracting these individual
background frames from the original sky frames to
generate more fine-scale sky frames.  These improved sky frames are again
median combined and added to the initial background, 
yielding a final sky frame for each exposure sequence.  
The high-order ($4\leq$$n$$\leq$$6$) distortion terms present in ISPI images 
were corrected using the IRAF \texttt{ccmap} and \texttt{mscimage} tasks 
by matching many (typically several hundred) well-detected sources to
2MASS.  Finally, a low-order background is fit to each sky-subtracted, distortion corrected image to remove any residual large-scale gradients.  

\begin{deluxetable*}{llcccc}
\tablecolumns{6}
\tablecaption{Log of ISPI Observations\label{logtab}}
\tablehead{ Date & Cluster & $N(J)$ & $N(K_{S})$ & Coadds($J$) & Coadds
  ($K_{S}$) }
\startdata
 13 Aug 2008 & NGC 104 & 4 & 12 & 12 & 6 \\
	     & NGC 6496 & 6 & 29 & 6 & 12,6 \\
 30 Sep 2009 & NGC 1851 & 4 & 29 & 3 & 6 \\
	     & NGC 288 & 10 & 26  & 3 & 6 \\
 01 Oct 2009 & NGC 362 & 15 & 34 & 3,6 & 6 \\
	     & NGC 1261 & 10  & 29 & 6 & 6 \\
	     & NGC 7099 & 15 & 34 & 3,6 & 6 \\
 28 Apr 2010 & NGC 2808 & 10 & 34 & 4 & 6 \\
	     & NGC 6304 & 10 & 35 & 4 & 6 \\
 29 Apr 2010 & NGC 4833 & 5 & 40 & 4 & 6 \\
 30 Apr 2010 & NGC 5927 & 10 & 33 & 4 & 6 \\
	     & NGC 6584 & 9 & 34 & 4 & 6 \\
\enddata
\end{deluxetable*}

\subsection{Photometry and Calibration}

\begin{figure}
\figurenum{1}
\includegraphics[width=0.5\textwidth]{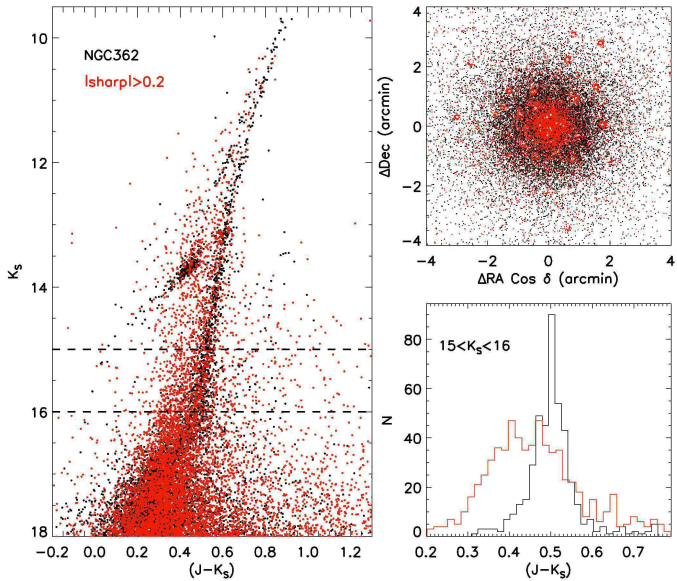}
\caption{Left: CMD of NGC 362 showing all stars which passed our photometric
  quality cuts (black) as well as those which failed the sharpness cut (red).
The dashed horizontal lines indicate the magnitude range used to construct a color
histogram of the lower red giant branch (lower right), illustrating the bias 
caused by stars
with higher absolute values of the sharpness parameter.  The
spatial distribution of the two samples relative to the cluster center (upper
right) illustrates the efficiency of the sharpness criterion to eliminate
spurious detections.}
\label{sharpfig}
\end{figure}

Point-spread function fitting (PSF) photometry was performed via 
iterative usage of the DAOPHOT/ALLFRAME suite
\citep{stet87,stet94} as described in \citet{mauropipe}.  To limit our
analysis to well-measured point sources, the resulting
instrumental catalogs were culled to retain only stars with a 
photometric error $\sigma$$(J-K_{S})\le$0.1 and 
$|$\texttt{sharp}$|\le$0.2.  This latter cut was found to
efficiently reject both spurious detections as well as stars with colors substantially affected
by crowding, illustrated in 
Fig.~\ref{sharpfig} for the example of NGC 362.  
There, detections which failed the
sharpness cut, plotted in red, are found not only to cluster around bright
sources spatially (as seen in the upper right panel), but also have RGB 
colors which are scattered preferentially blueward due to blending, seen
in the color-magnitude diagram (CMD) as well as a color histogram of the lower RGB (bottom right panel of Fig.~\ref{sharpfig}).
This effect is discussed, for example, by \citet{stetcrowd}, 
and demonstrates why a fairly
stringent sharpness cut is needed despite the loss of some real
sources in order to guard against systematic color offsets
as a result of photometric blends.  We caution that by electing
straightforward self-consistent photometric quality cuts, a small fraction 
of spurious detections may remain in our final catalogs, although this has a
negligible impact on our results since our analyses typically employ stars well
above our detection limit, and any remaining spurious detections likely fall well redward of the cluster sequences (see Fig.~\ref{sharpfig}). 

\begin{figure*}
\figurenum{2}
\includegraphics[width=0.49\textwidth]{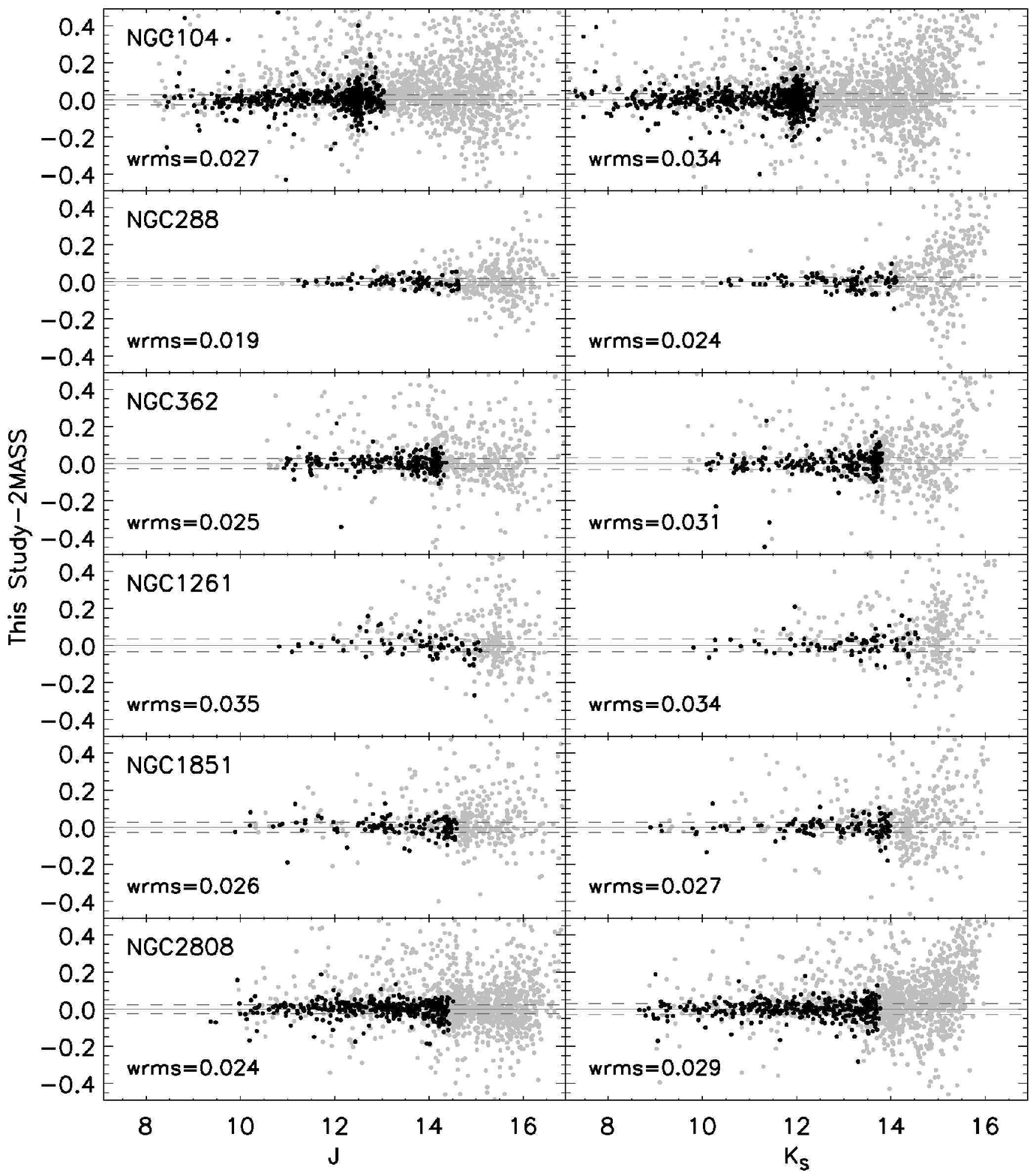}
\includegraphics[width=0.49\textwidth]{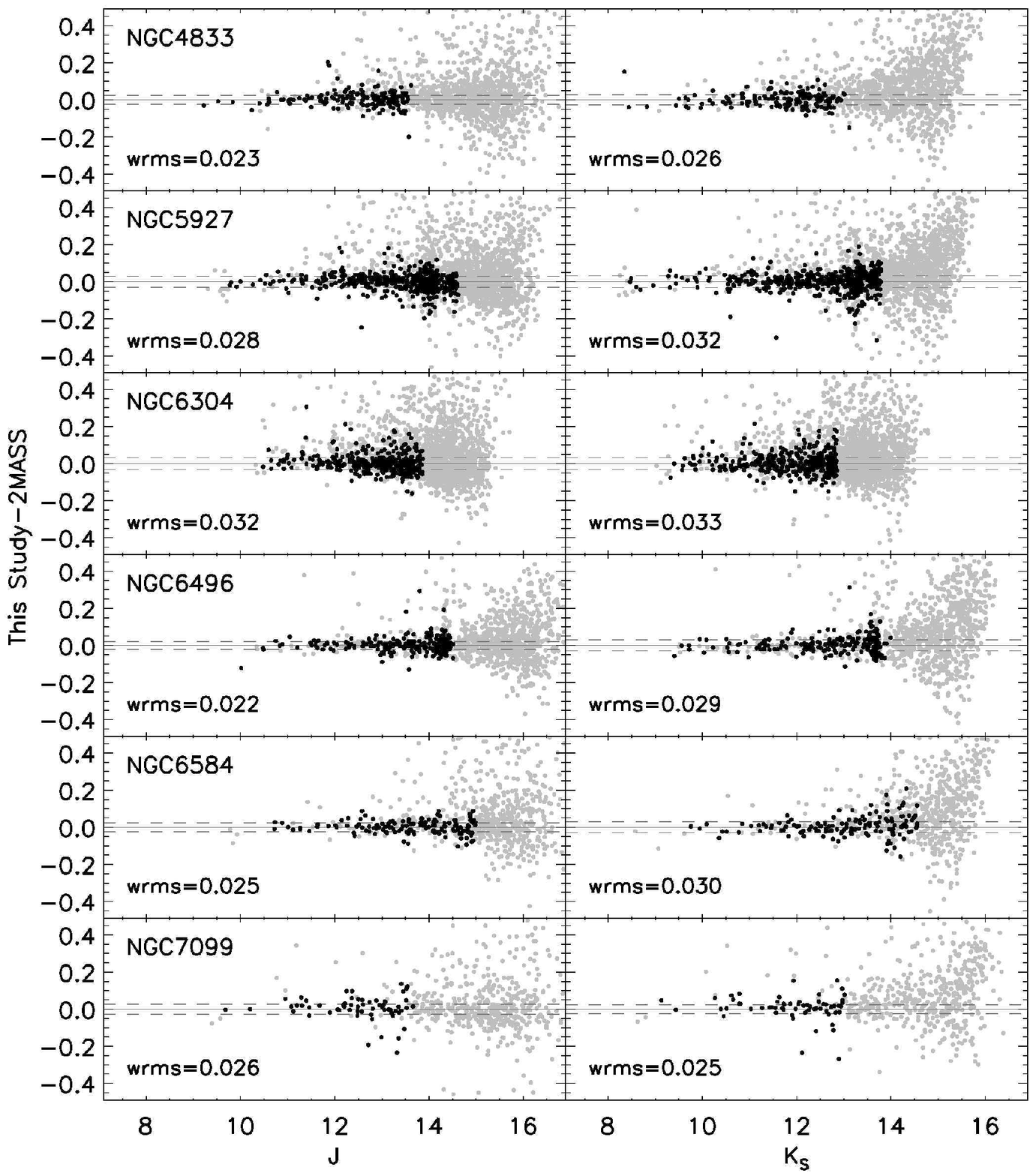}
\caption{A comparison of our calibrated photometry in the $J$ (left) and
  $K_{S}$ (right) filters with photometry from the 2MASS PSC.  All matched
  stars are shown in grey, and 2MASS stars used for photometric calibration
  are overplotted in black.  In each plot, the solid grey line represents
  equality while the dashed grey lines represent the weighted RMS deviation
  among the calibrators.}
\label{comp2massfig}
\end{figure*}

In order to
facilitate comparison and concatenation with existing GGC near-IR photometry
(see Sect.~\ref{compphotsect}) as well as direct comparison to multiple sets of 
evolutionary models (see Sect.~\ref{modelsect}), 
we have chosen to calibrate our instrumental
catalogs to the 2MASS photometric system.   
Stars astrometrically matched to 2MASS were only used as
photometric calibrators if they satisfied a more stringent set of criteria:
They must be matched to a 2MASS point source to
within the astrometric rms ($<$0.2$\arcsec$), isolated 
(lacking neighbors within 4 mag inside a 2.5$\arcsec$ radius, corresponding to
a contaminating flux of $<$2.5\%) and be bright enough to remain unaffected by
crowding in 2MASS (see below).    
To calculate transformations from the instrumental magnitudes in our
PSF catalogs to the 2MASS photometric system, we employed 
classical linear transformation equations of the form $m-M=a+b(J-K_{S})$,
where $m$ and $M$ denote instrumental and standard magnitudes respectively.   
We solve for the coefficients $a$ and $b$ using least squares fitting, but
employing weighting factors to downweight discrepant data points in lieu of
a sigma clip\footnote{The algorithm is based on a series of five lectures
  presented at "V Escola Avancada de Astrofisica" by Peter Stetson
  \url{http://ned.ipac.caltech.edu/level5/Stetson/Stetson\_contents.html}, \url{http://www.cadc.hia.nrc.gc.ca/community/STETSON/\\homogenous/Techniques/}}
 \citep{mauropipe,cohen6544}, and we apply these same weighting factors to
calculate the weighted root mean square deviation (wrms) of the residuals.
A comparison between the results of our calibration procedure and the
magnitudes reported in the 2MASS PSC are shown in Fig.~\ref{comp2massfig},
with the wrms given in each plot.  The asymmetry seen at the faint end 
among stars positionally matched to 2MASS is due to the brighter 
completeness limit and
lower spatial resolution of 2MASS, resulting in a systematic brightward
deviation in 2MASS magnitudes.  This effect serves as a cautionary note
against calibrating to 2MASS in cases where only the faintest 2MASS stars are
unsaturated, and for this reason we exclude faint 2MASS sources ($K_{S}\gtrsim$14 depending on stellar density) 
from use as photometric calibrators (see Fig.~\ref{comp2massfig}).  
Statistics regarding calibration
uncertainties for each cluster, including the wrms values, 
photometric zero point uncertainties, and the number of stars from 2MASS used
for photometric calibration, are summarized in Table
\ref{2masscaltab}.  As our photometry saturates slightly below
the tip of the cluster red giant branch in some cases, 
we have supplemented our catalogs with 2MASS photometry for bright stars 
which are absent due to saturation.

\begin{deluxetable}{lccccr}
\tablecolumns{6}
\tablewidth{21pc}
\tablecaption{Photometric Calibration Uncertainties with Respect to 2MASS}
\tablehead{ Cluster & \multicolumn{2}{c}{wrms} & \multicolumn{2}{c}{$\sigma(zpt)$} &
N } 
\startdata
 & $J$ & $K_{S}$ & $J$ & $K_{S}$ & \\
\hline
NGC 104 & 0.027 & 0.034 & 0.005 & 0.006 & 647 \\
NGC 288 & 0.019 & 0.024 & 0.013 & 0.017 & 89 \\
NGC 362 & 0.025 & 0.031 & 0.007 & 0.010 & 222 \\
NGC 1261 & 0.035 & 0.034 & 0.014 & 0.014 & 76 \\
NGC 1851 & 0.026 & 0.027 & 0.010 & 0.011 & 118 \\
NGC 2808 & 0.024 & 0.029 & 0.006 & 0.008 & 347 \\
NGC 4833 & 0.023 & 0.026 & 0.009 & 0.010 & 164 \\
NGC 5927 & 0.028 & 0.032 & 0.005 & 0.006 & 495 \\
NGC 6304 & 0.032 & 0.033 & 0.010 & 0.011 & 360 \\
NGC 6496 & 0.022 & 0.029 & 0.006 & 0.007 & 220 \\
NGC 6584 & 0.025 & 0.030 & 0.007 & 0.009 & 135 \\
NGC 7099 & 0.026 & 0.025 & 0.033 & 0.036 & 61 \\ 
\enddata
\label{2masscaltab}
\end{deluxetable}

\subsection{Comparison to Previous Photometry\label{compphotsect}}

\begin{figure}
\figurenum{3}
\includegraphics[width=0.48\textwidth]{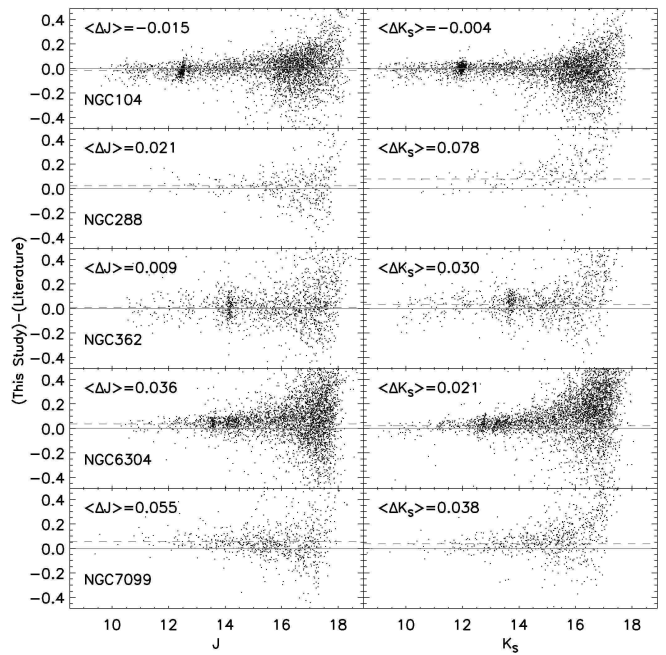}
\caption{Comparison of our 2MASS-calibrated $JK_{S}$ photometry with that of
  \citet{salaris47tuc} for NGC 104 (top panel) 
  and \citet{v04obs} and \citet{vof05} for the remainder of the clusters
  illustrated.  The solid grey 
line represents equality, and the dotted grey line represents the mean offset,
calculated using a 2.5$\sigma$ clip in bins of 1 mag and given in the upper
left of each panel.}
\label{compphotv04}
\end{figure}

Four of our target clusters have been observed by 
\citet{v04obs,vof05}\footnote{Also see the Bulge Globular Cluster Archive
  at \url{http://www.bo.astro.it/$\tilde{ }$GC/ir\_archive}}, and near-IR
photometry of 47 Tuc has been presented by \citet{salaris47tuc},
allowing a direct comparison of our 2MASS-calibrated $JK_{S}$ photometry to
theirs.    
In all cases, we
recover the majority of previously detected sources, 
and a cluster-by-cluster comparison of
magnitude difference as a function of magnitude is shown in 
Fig.~\ref{compphotv04}.  There, we have computed the
mean magnitude difference for each cluster in each filter using a weighted 
2.5$\sigma$ clip in magnitude bins while excluding stars
faintward of the observed luminosity function peak.  Our calibrated photometry
is in agreement with earlier studies in light of total calibration uncertainties,
and although
offsets of $>$0.05 mag are seen for NGC 288 in the $K_{S}$ band (and to a
lesser extent NGC 7099 in $J$), the
direct comparison to 2MASS in Fig.~\ref{comp2massfig} gives no reason to be
doubtful about the calibration of these clusters given our calibration
uncertainties listed in Table \ref{2masscaltab} and the zeropoint uncertainty
of $\pm$0.05 mag cited by \citet{vof05}.

\begin{figure}
\figurenum{4}
\epsscale{0.5}
\includegraphics[width=0.49\textwidth]{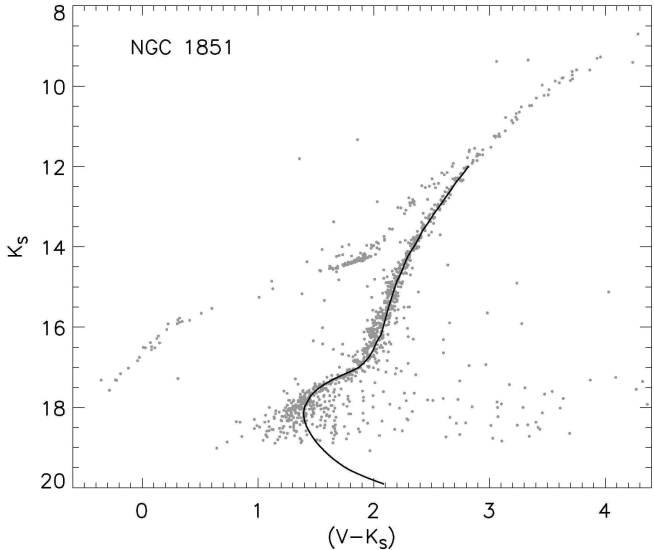}
\caption{Observed $K_{S},(V-K_{S})$ CMD of NGC 1851, obtained by matching our
  ISPI photometry with publicly available optical photometry (see
  text for details).  The fiducial sequence of \citet{brasseur} is
  overplotted as a black solid line.}
\label{compbrasseur}
\end{figure}

In addition, 
there is one cluster in our sample (NGC 1851) in common with the study
of \citet{brasseur}.  To compare their $(V-K_{S})$ fiducial sequence with our
observations, we have matched our photometry to publicly available optical
photometry from the archive of P.~B.~Stetson\footnote{See
  \url{http://www3.cadc-ccda.hia-iha.nrc-cnrc.gc.ca/en/community/STETSON/}} as
well as the photometric catalog available from the ACS GGC Treasury Survey
\citep{sarajedini07}.  In the latter case, the offsets listed by
\citet{marenacs} were applied to the \citet{sirianni} transformed magnitudes
to place them on the photometric system employed by Stetson.  As the
ACS catalogs are astrometrically calibrated to 2MASS
\citep{andersonacs}, a simple matching algorithm with an initial tolerance of
0.6$\arcsec$ and an iterative 5$\sigma$ clip functions well, and we
recovered 99\% of ISPI sources in the ACS field of view with an astrometric
rms of 0.134$\arcsec$.  
The resulting matched CMD is shown in Fig.~\ref{compbrasseur} 
with the optical-IR fiducial sequence of \citet{brasseur} overplotted.

\section{Color-Magnitude Diagrams and Fiducial Sequences}

\subsection{Construction of Fiducial Sequences\label{fidsect}}

To construct fiducial sequences representative of the RGB 
in each cluster, we fit a low-order ($n\le$3)
polynomial in the $(J-K_{S})$ vs. $K_{S}$ plane in a similar manner as previous studies \citep[e.g.][]{ferraro99,ferraro00,cohen6544}: 
First, a rough color-magnitude cut is made in the cluster CMD to
select the region of the RGB.  Next, the RGB is divided into bins of 0.5 mag, 
and the median color and magnitude (and their uncertainty) are calculated in 
each bin.  A polynomial is then
fit to these median points, iteratively rejecting stars more than 2$\sigma$
from the polynomial in each bin 
until convergence is indicated by the total number of stars 
remaining constant to within 2\% since the previous iteration.

\begin{figure*}
\centering
\figurenum{5}
\includegraphics[width=0.81\textwidth]{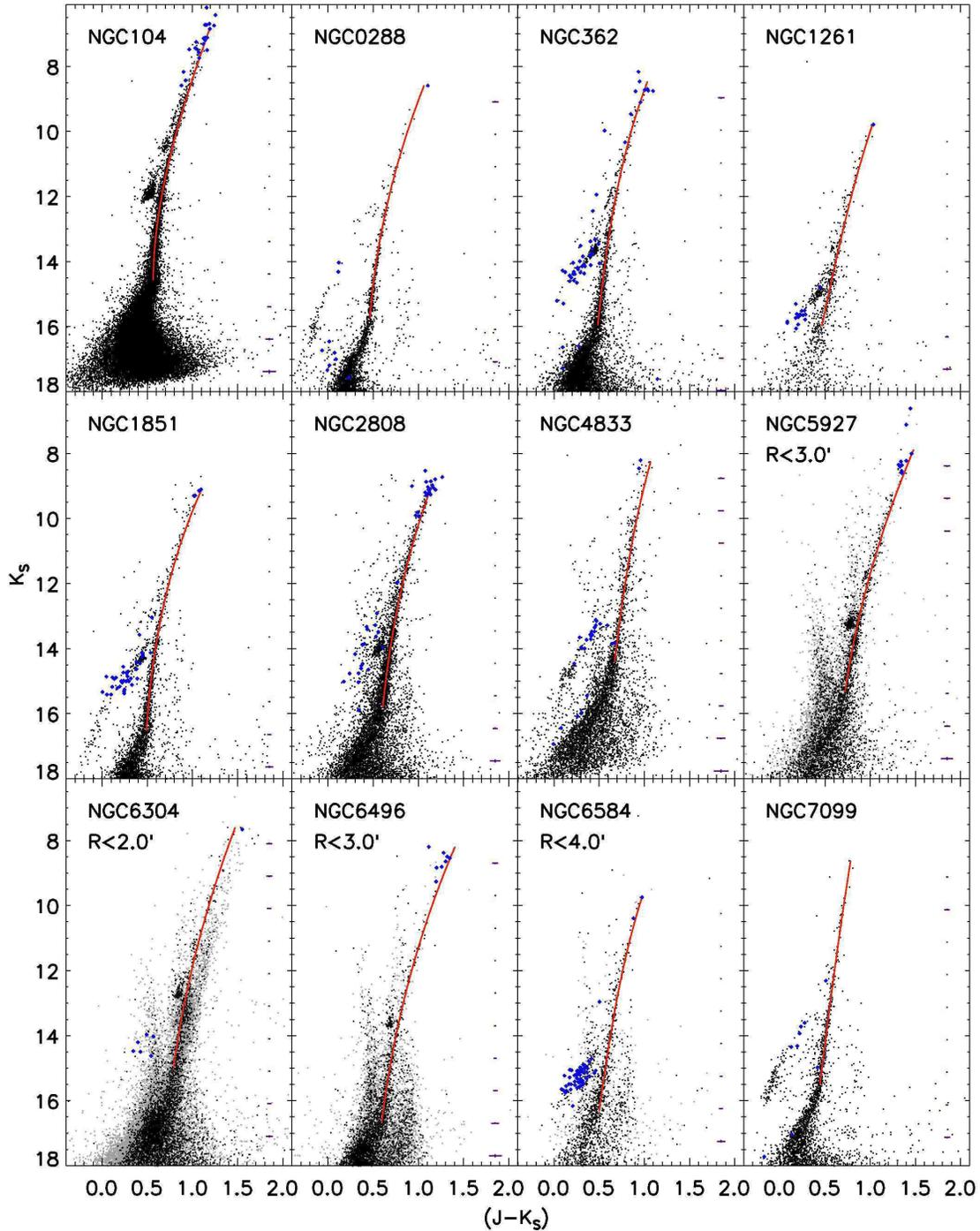}
\caption{CMDs of our target clusters, with fiducial sequences overplotted in
  red and known variables as blue diamonds.  In cases where a radial cut was
  necessary to reduce field contamination, the location of the radial cut in
  arcmin from the cluster center is given, and stars outside this radius are
  plotted in grey rather than black.  Median color and magnitude errors in
  bins of 1 mag are shown along the right side of each CMD.}
\label{cmdsjk}
\end{figure*}

In Fig.~\ref{cmdsjk} we present $J-K_{S},K_{S}$ CMDs of all of our target
clusters, with the RGB fiducial sequences overplotted in red and
median photometric errors in bins of 1 mag 
indicated along the right-hand side of each CMD.  
All varibles matched to our
photometry from the most recent version of the \citet{clement} catalog of
variable stars in GGCs\footnote{See
  \url{http://www.astro.utoronto.ca/~cclement/cat/listngc.html}} are
overplotted as blue diamonds and excluded from further analyses.

Our photometric catalogs are available electronically,
and their format is illustrated in Table 
\ref{catexampletab}.

\begin{deluxetable}{cccccc}
\tablecolumns{6}
\tablecaption{NGC 104 Photometric Catalog}
\tablehead{ RA (J2000) & Dec (J2000) & $J$ & $\sigma$$J$ & $K_{S}$ &
  $\sigma$$K_{S}$ }
\startdata
    6.283727 & -72.069279 & 10.9506 & 0.0263 & 10.3440 & 0.0044 \\
    6.283700 & -72.031622 & 14.2151 & 0.0155 & 13.7408 & 0.0075 \\
    6.283753 & -72.133077 & 15.2417 & 0.0217 & 14.9558 & 0.0144 \\
    6.283345 & -72.093668 & 16.7742 & 0.0621 & 16.5855 & 0.0298 \\
    6.283288 & -72.071959 & 17.8108 & 0.0765 & 17.2019 & 0.0482 \\
    6.283036 & -72.096528 & 17.0105 & 0.0611 & 16.9107 & 0.0429 \\
    6.282974 & -72.112954 & 16.2559 & 0.0342 & 16.1854 & 0.0254 \\
    6.282944 & -72.063674 & 11.2960 & 0.0234 & 10.6157 & 0.0042 \\
    6.282786 & -72.060823 & 16.6431 & 0.0446 & 16.2261 & 0.0231 \\
    6.282719 & -72.022186 & 17.9398 & 0.0723 & 17.4222 & 0.0490 \\
\enddata
\label{catexampletab}
\tablecomments{The full photometric catalogs for all target clusters are
  available electronically; a portion is shown here to illustrate their form and content}
\end{deluxetable}

\subsection{The Absolute Magnitude Plane: Cluster Distance, Reddening and Metallicity\label{calibclustersect}}

One of the primary goals of the present study is to use a self-consistent set 
of cluster distances, reddenings and metallicities to compare observed 
evolutionary sequences to models.  Furthermore, with an eye towards heavily
reddened stellar systems for which the use of optical photometry to measure
cluster parameters is prohibitive, 
we also rederive empirical relations
describing the RGB shape as a function of metallicity.    
To this end, we select distances and 
reddenings from \citet{d10} and $[Fe/H]$ (and uncertainties) from \citet{c09}.
For consistency with \citet{d10} we employ the total to selective extinction 
ratios $R_{V}$ given by \citet{sirianni} for a G2 star, and values of
$A_{J}/E(B-V)$=0.899 and $A_{Ks}/E(B-V)$=0.366 from \citet{casagrande}.  For the two GGCs excluded from \citet{d10} due to the
presence of multiple stellar populations (NGC 1851, NGC 2808), we employ parameters obtained identically as for the remainder of the \citet{d10} sample (A. Dotter, private communication, also see \citealt{hbparams1}).

\begin{figure}
\figurenum{6}
\label{fidsabs}
\includegraphics[width=0.45\textwidth]{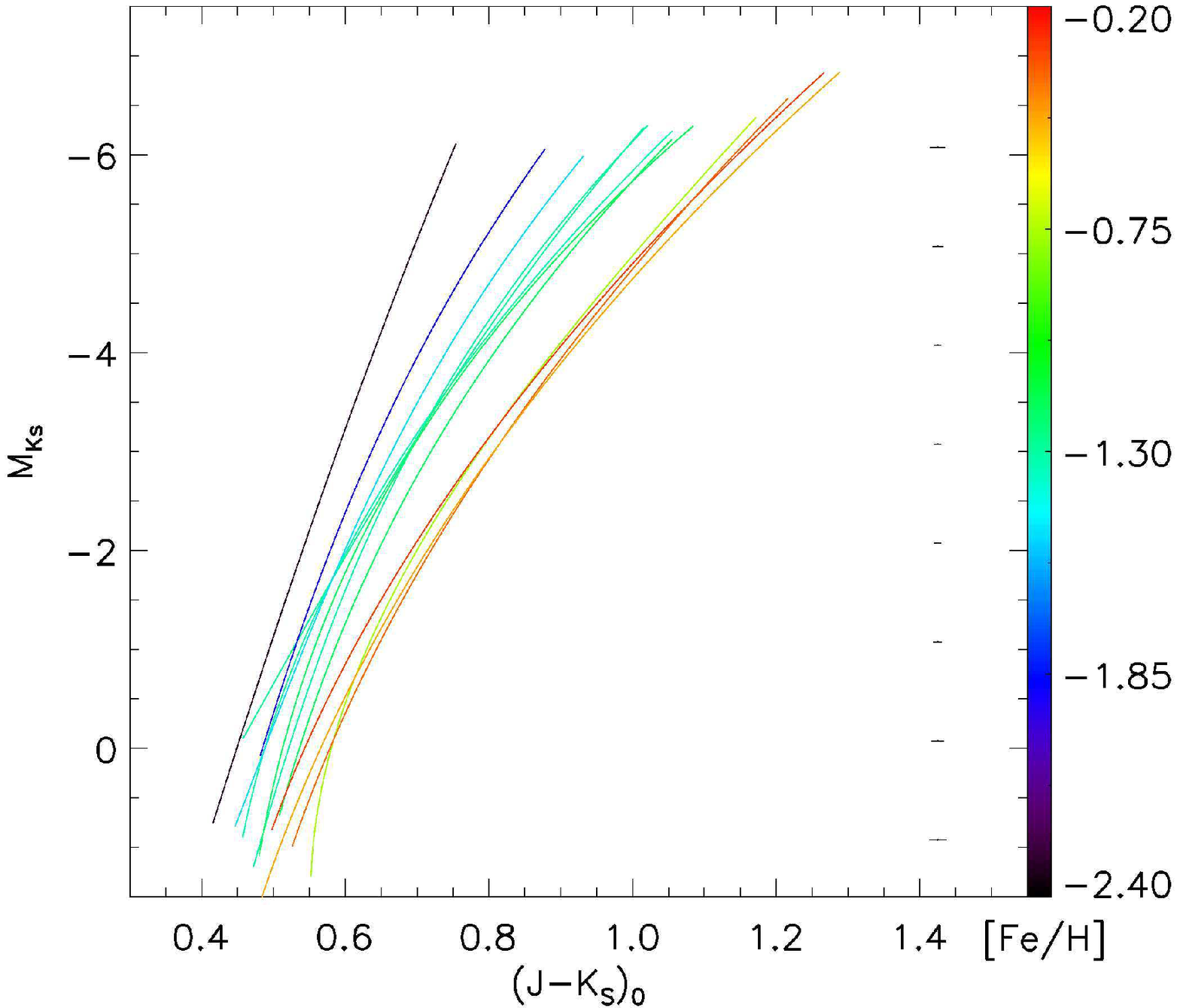}
\caption{Our fiducial sequences in the $M_{K},(J-K_{S})_{0}$ 
plane, color coded by $[Fe/H]$ value.  Median photometric errors in magnitude
bins are shown along the right-hand side.}
\end{figure}

Our adopted values for the cluster distance moduli, reddening and metallicity
are listed in Table \ref{clusparamtab}.  In light of cluster-to-cluster
variations in $\alpha$-enhancement, we also calculate the global metallicity
$[M/H]=[Fe/H]+Log(0.638f_{\alpha}+0.362)$ where $f_{\alpha}=10^{[\alpha/Fe]}$  \citep{salaris}.   Observed values of $[\alpha/Fe]$ from
\citet{c10} are listed in Table \ref{clusparamtab} where available, as well as
the values we assume for our isochrone comparison (discussed below in
Sect.~\ref{modelsect}).  We also list the corresponding global metallicity $[M/H]$, including its uncertainty calculated following \citet[][see their eq.~7]{natafbump} and conservatively assuming $\sigma$$[\alpha/Fe]$=0.1. 
In Fig.~\ref{fidsabs},
we display our fiducial sequences, shifted to the dereddened plane 
and color coded by $[Fe/H]$.  

\begin{deluxetable*}{lccccccccc}
\tablecolumns{9}
\tablewidth{40pc}
\tablecaption{Cluster Parameters}
\tablehead{Cluster & RA (J2000)\tablenotemark{a} & Dec (J2000)\tablenotemark{a} & $(m-M)_{0}$\tablenotemark{b} & $E(B-V)$\tablenotemark{b} & $[Fe/H]$\tablenotemark{c} & $[\alpha/Fe]$\tablenotemark{d} & $[\alpha/Fe]$ & $[M/H]$\tablenotemark{e} & $\Delta$$(V-I)$\tablenotemark{b}  \\ & & & & & & observed & assumed & assumed & } 
\startdata
NGC104 & 00:24:05.71 & -72:04:52.7 & 13.26 & 0.023 & -0.76$\pm$0.02 & 0.42 & 0.4$\pm$0.1 & -0.47$\pm$0.08 & 0.153$\pm$0.003\\
NGC288 & 00:52:45.24 & -26:34:57.4 & 14.83 & 0.013 & -1.32$\pm$0.02 & 0.42 & 0.4$\pm$0.1 & -1.03$\pm$0.08 & 1.022$\pm$0.025 \\
NGC362 & 01:03:14.26 & -70:50:55.6 & 14.76 & 0.023 & -1.30$\pm$0.04 & 0.30 & 0.4$\pm$0.1 & -1.01$\pm$0.09 & 0.195$\pm$0.003 \\
NGC1261 & 03:12:16.21 & -55:12:58.4 & 16.08 & 0.013 & -1.27$\pm$0.08 & & 0.4$\pm$0.1 & -0.98$\pm$0.11 & 0.201$\pm$0.005 \\
NGC1851 & 05:14:06.76 & -40:02:47.6 & 15.42 & 0.020 & -1.18$\pm$0.08 & 0.38 & 0.4$\pm$0.1 & -0.89$\pm$0.11 & 0.234$\pm$0.011 \\
NGC2808 & 09:12:03.10 & -64:51:48.6 & 15.05 & 0.183 & -1.18$\pm$0.04 & 0.33 & 0.4$\pm$0.1 & -0.89$\pm$0.09 & 0.966$\pm$0.025 \\
NGC4833 & 12:59:33.92 & -70:52:35.4 & 14.19 & 0.359 & -1.89$\pm$0.05 & & 0.4$\pm$0.1 & -1.60$\pm$0.10 & 0.900$\pm$0.029 \\
NGC5927 & 15:28:00.69 & -50:40:22.9 & 14.57 & 0.399 & -0.29$\pm$0.07 & & 0.2$\pm$0.1 & -0.15$\pm$0.10 & 0.112$\pm$0.007 \\
NGC6304 & 17:14:32.25 & -29:27:43.3 & 13.99 & 0.481 & -0.37$\pm$0.07 & & 0.2$\pm$0.1 & -0.23$\pm$0.10 & 0.105$\pm$0.004\\
NGC6496 & 17:59:03.68 & -44:15:57.4 & 14.95 & 0.216 & -0.46$\pm$0.07 & & 0.2$\pm$0.1 & -0.32$\pm$0.10 & 0.107$\pm$0.008 \\
NGC6584 & 18:18:37.60 & -52:12:56.8 & 15.71 & 0.079 & -1.50$\pm$0.09 & & 0.4$\pm$0.1 & -1.21$\pm$0.12 & 0.408$\pm$0.062 \\
NGC7099 & 21:40:22.12 & -23:10:47.5 & 14.72 & 0.054 & -2.33$\pm$0.02 & 0.37 & 0.4$\pm$0.1 & -2.04$\pm$0.08 & 0.872$\pm$0.006 \\
\enddata
\tablenotetext{a}{From \citet{goldsbury}}
\tablenotetext{b}{From \citet{d10}}
\tablenotetext{c}{From \citet{c09}}
\tablenotetext{d}{From \citet{c10}}
\tablenotetext{e}{Calculated using $[M/H]=[Fe/H]+Log(0.638f_{\alpha}+0.362)$, where $[\alpha/Fe]$=$[\alpha/Fe]$(assumed)}
\label{clusparamtab}
\end{deluxetable*}

\section{Models Versus Data on the Red Giant Branch}

\subsection{Adopted Models\label{modelsect}}

Our observations are compared to the following sets of evolutionary
models, all of which have been updated recently, including 
the capability to generate $\alpha$-enhanced isochrones:

\noindent1. Dartmouth Stellar Evolutionary Database (DSED; \citealt{dotter08})
isochrones with $[\alpha/Fe]$=0.4.

\noindent2. Victoria-Regina (VR; \citealt{v14}) isochrones with $[\alpha/Fe]$=0.4. 

\noindent3. Bag of Stellar Tracks and Isochrones (BaSTI; \citealt{basti}) canonical $\alpha$-enhanced isochrones.

\noindent4. Yale-Yonsei (YY; \citealt{yy1,yy2}) isochrones with $[\alpha/Fe]$=0.4. 

\noindent5. Princeton-Goddard-PUC (PGPUC; \citealt{pgpuc}) isochrones, 
with $[\alpha/Fe]$=0.3 (this is the
highest degree of $\alpha$ enhancement available over GGC parameter space).

Any critical attempt to evaluate these models versus each other on equal 
footing is not truly 
an apples-to-apples comparison, as they all
differ in various input assumptions
(e.g.~solar abundances, $\alpha$-enhanced chemical
compositions,   
treatment of convection and atomic diffusion, color-$T_{eff}$ relations).
However, a direct comparison between different near-IR isochrones has yet to be
published, and comparisons between individual isochrones and high-quality data
are scarce.  Here we take advantage of our homogenous photometry of 12 GGCs
covering a range of distance, reddening and metallicity to objectively
explore how well existing isochrones fit observational data in the near-IR. 
That being said, in a few cases some modifications were
made to the available default settings of various models in an attempt to
maximize consistency.  First, the near-IR colors and magnitudes output by the BaSTI and YY models are on the
Johnson-Cousins-Glass photometric system \citep{bb}, so they were converted to
the 2MASS photometric system using the relations of \citet{carpenter}.
Second, the VR and PGPUC models, rather than assuming a $\Delta$$Y/\Delta$$Z$
relation (or having one optionally available), give the user the ability to
interpolate simultaneously in $Y$, $[\alpha/Fe]$ and $[Fe/H]$ (or $Z$).  In these
cases, we have chosen values of $Y$ in order to be consistent with \citet{d10}, 
increasing $Y$ slightly with $[Fe/H]$ to approximate a $\Delta$$Y/\Delta$$Z$=1.4 relation.  In any case, 
minor model-to-model differences in $\Delta$$Y$ are inconsequential to our results
since the location of the RGB in the near-IR is quite insensitive to He variations
(see Sect.~\ref{hesect}) 

\subsection{Cluster by Cluster Comparison\label{isocompsect}}

\begin{figure*}[t!]
\centering
\figurenum{7}
\includegraphics[width=0.35\textwidth]{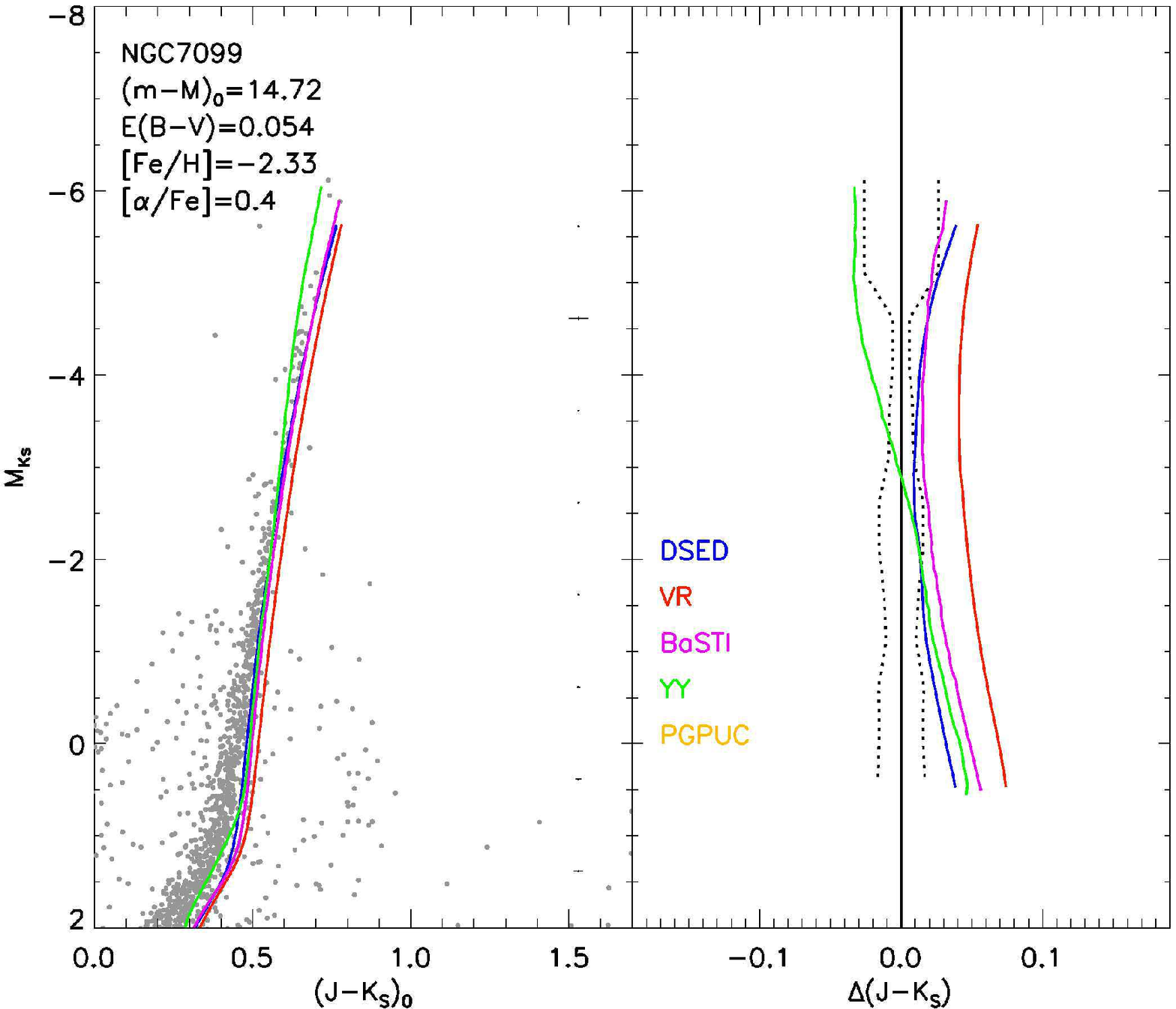}
\includegraphics[width=0.35\textwidth]{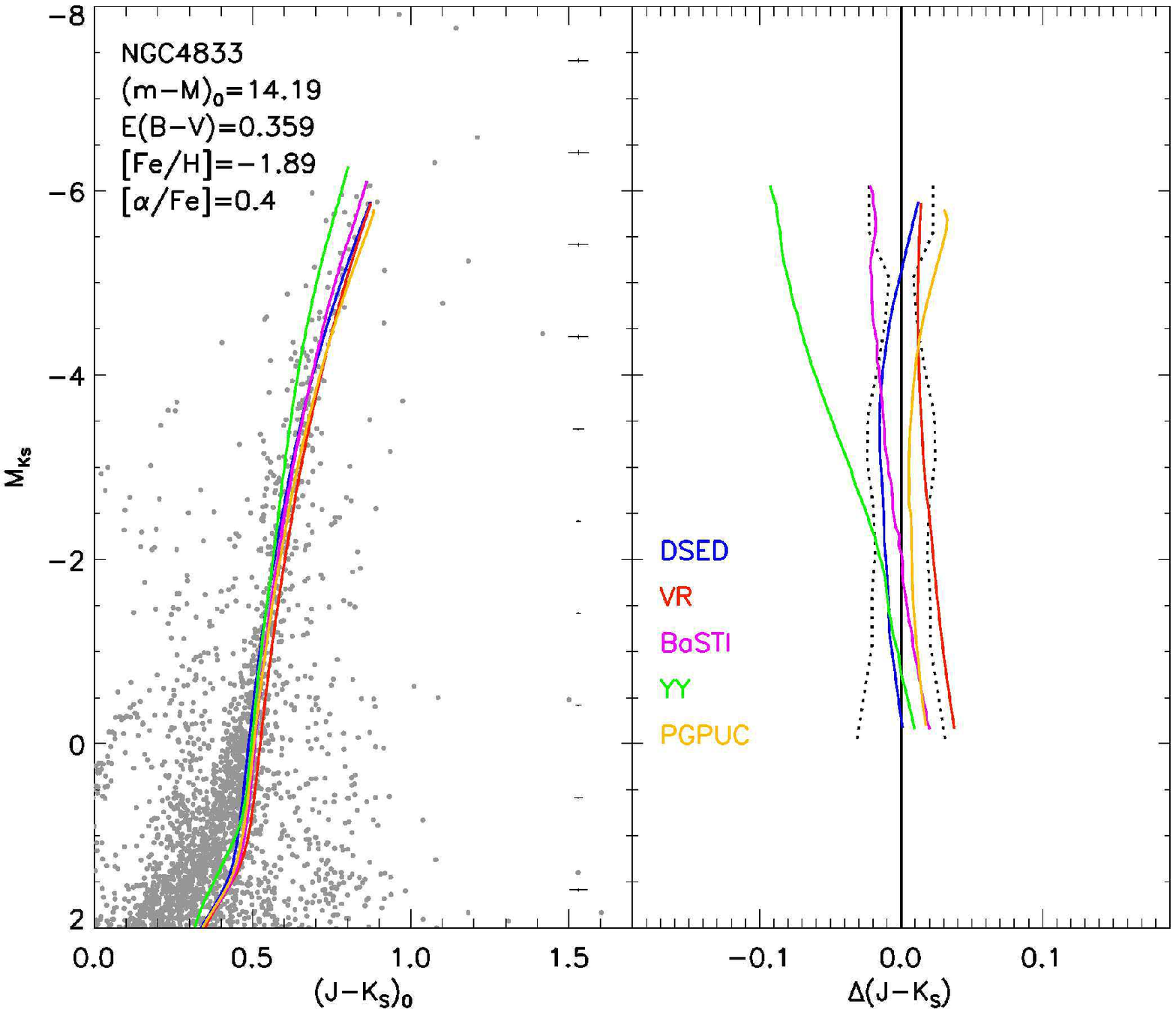}\\
\includegraphics[width=0.35\textwidth]{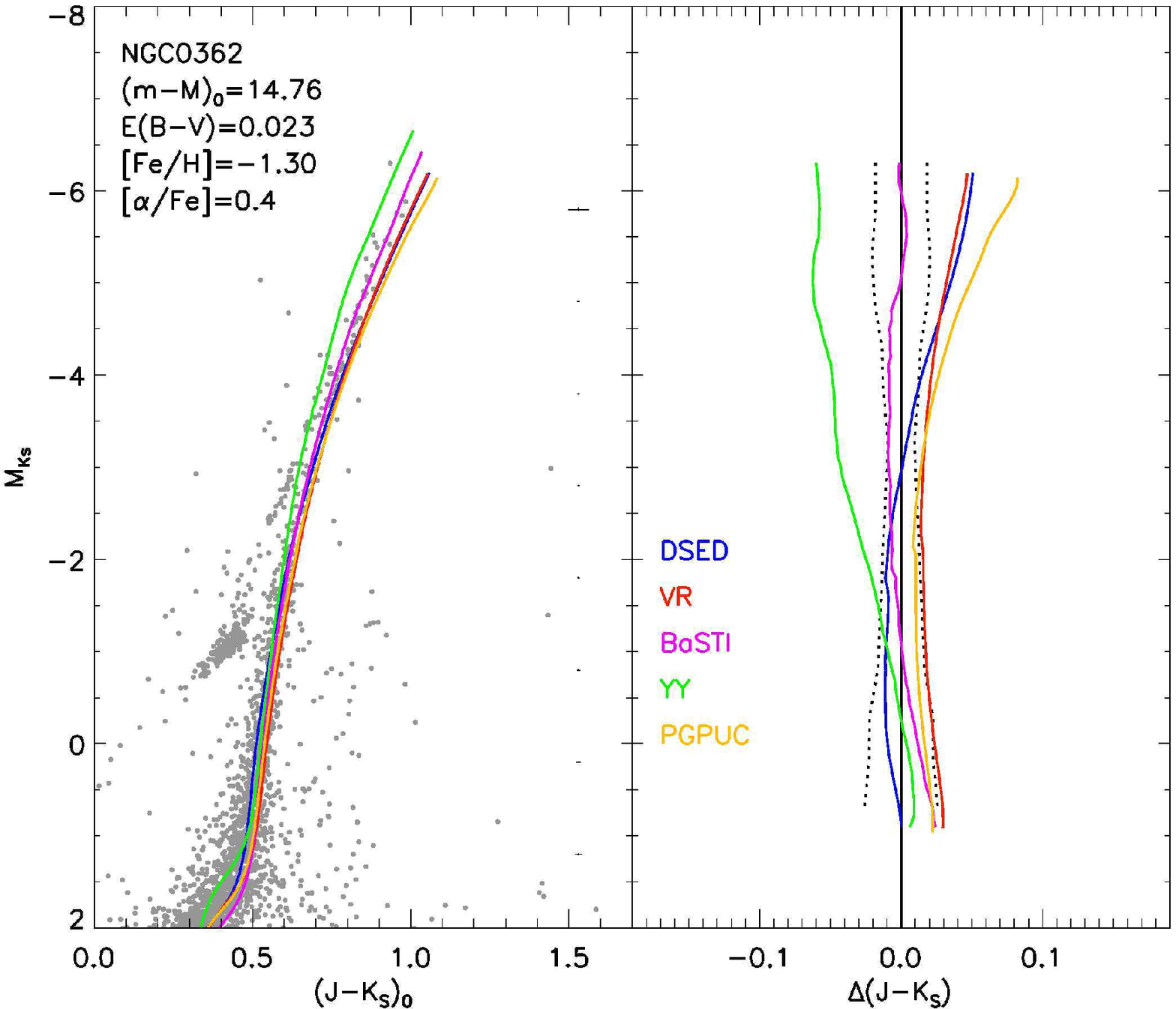}
\includegraphics[width=0.35\textwidth]{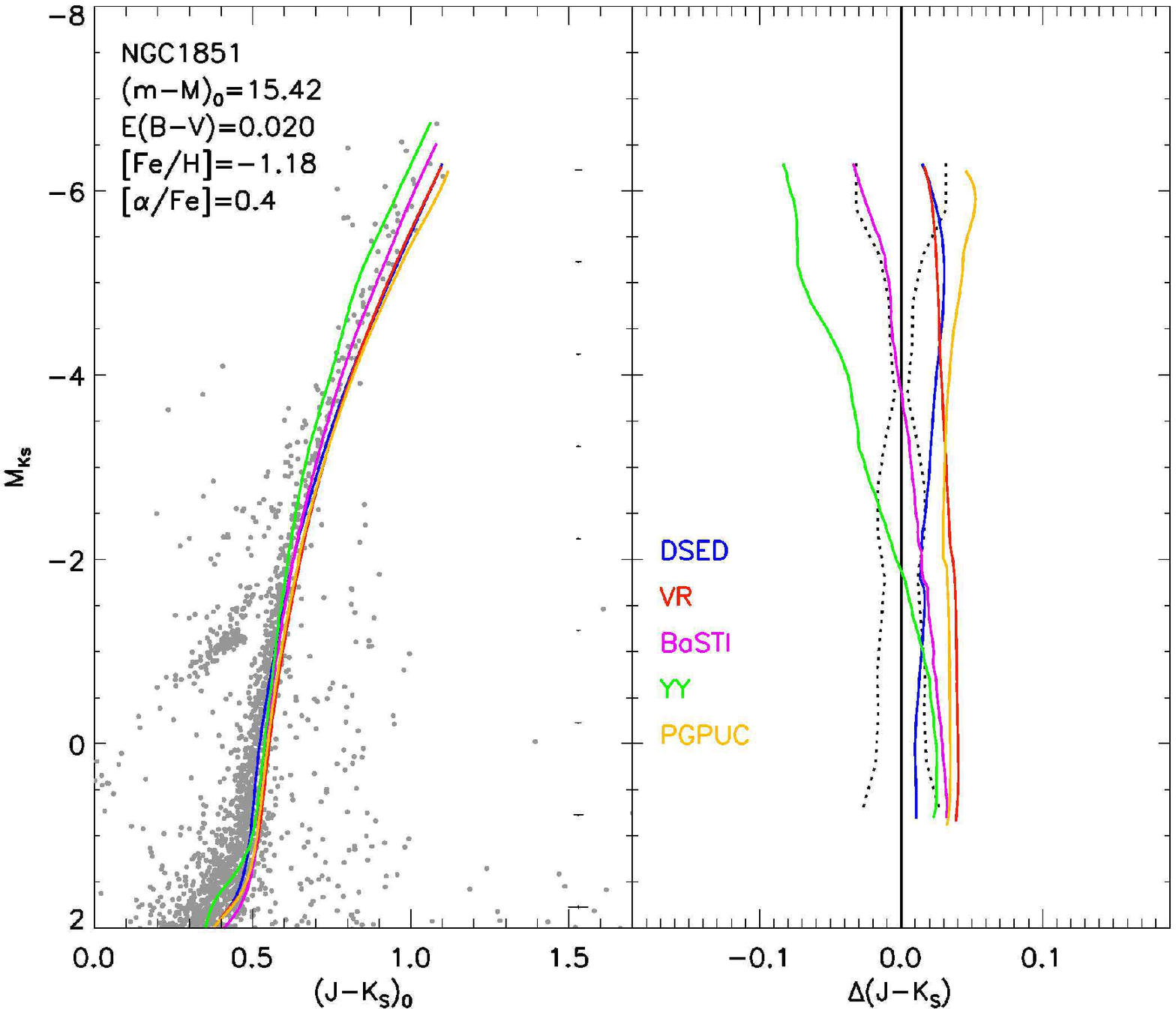}\\
\includegraphics[width=0.35\textwidth]{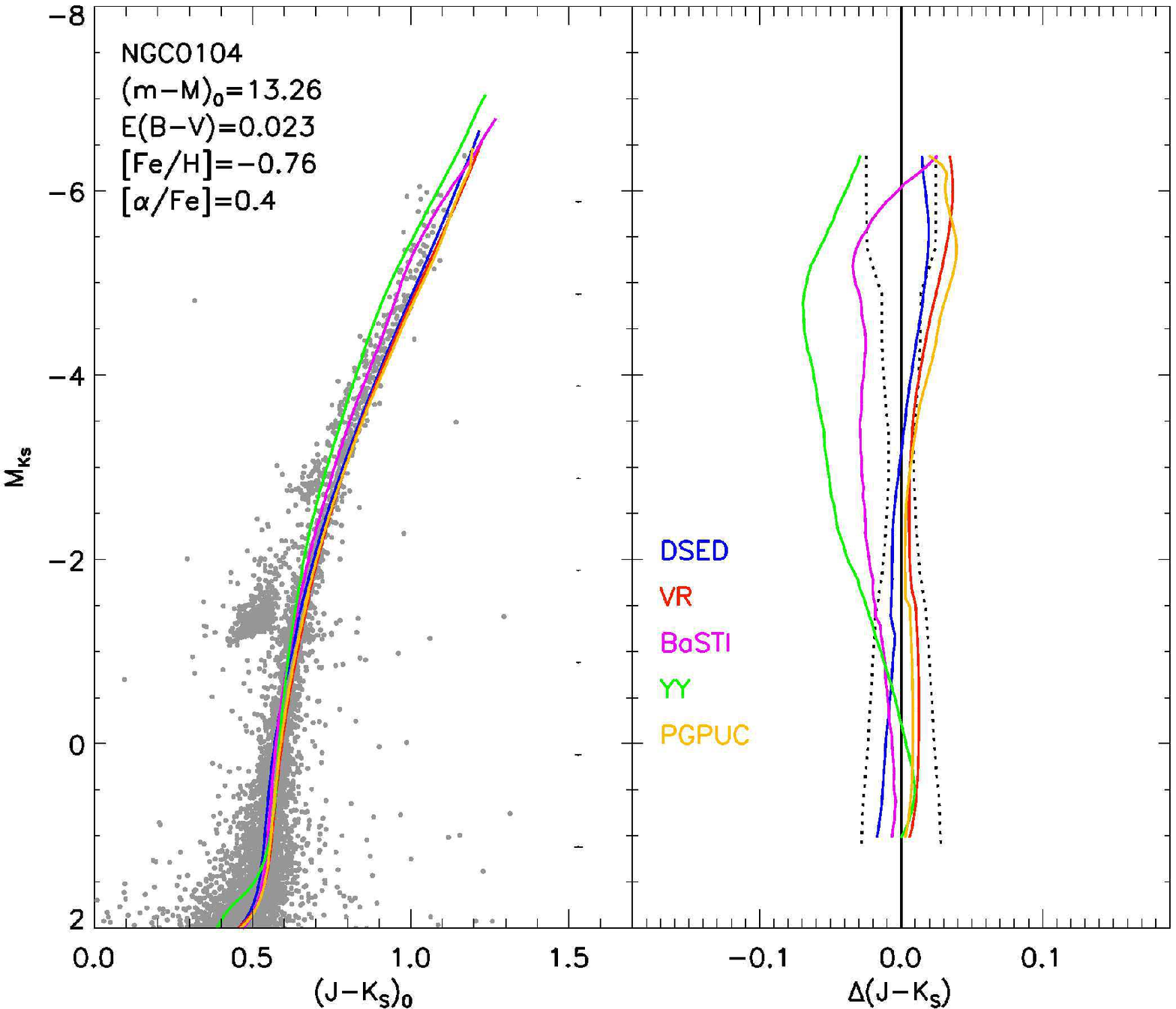}
\includegraphics[width=0.35\textwidth]{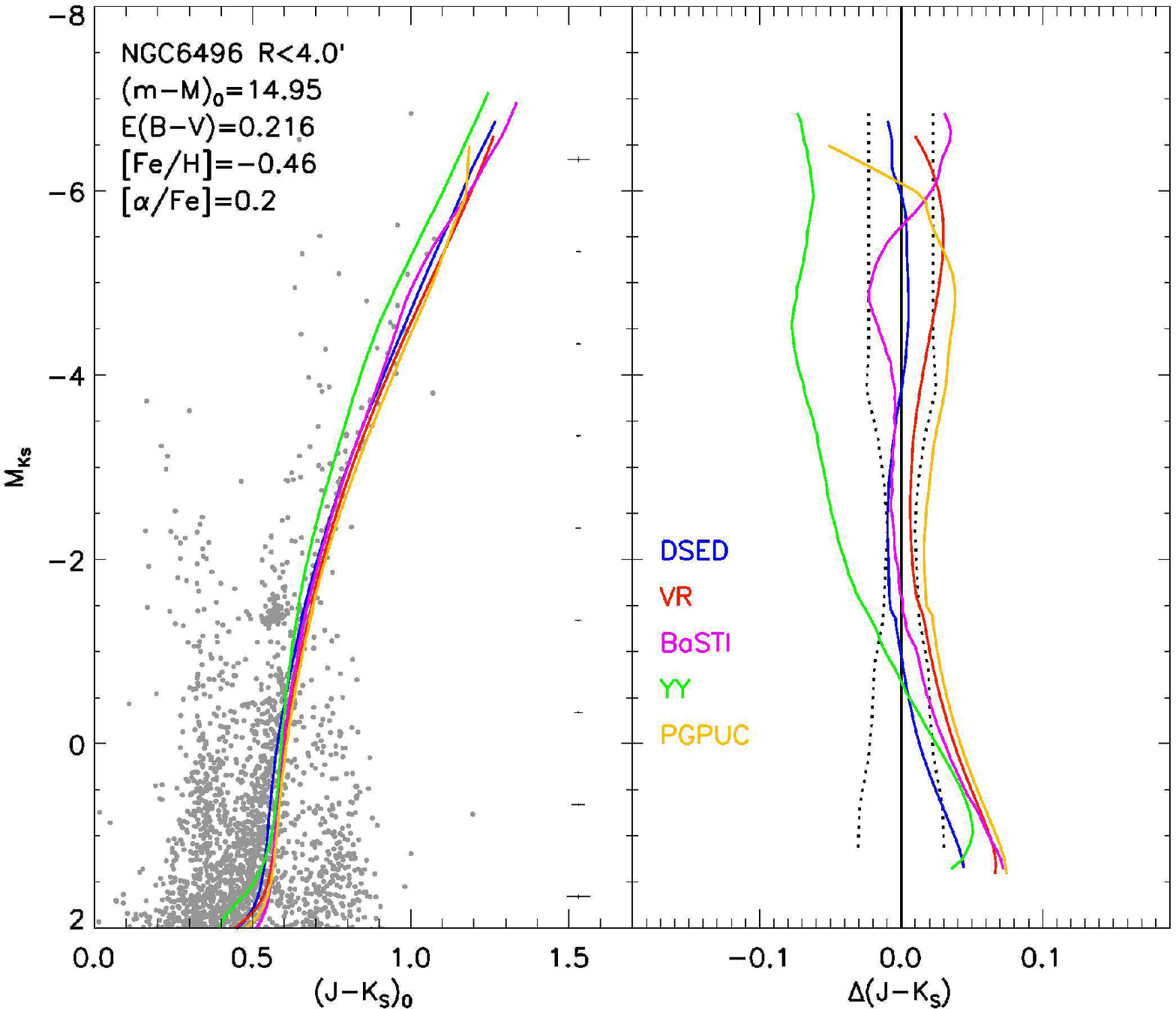}\\
\includegraphics[width=0.35\textwidth]{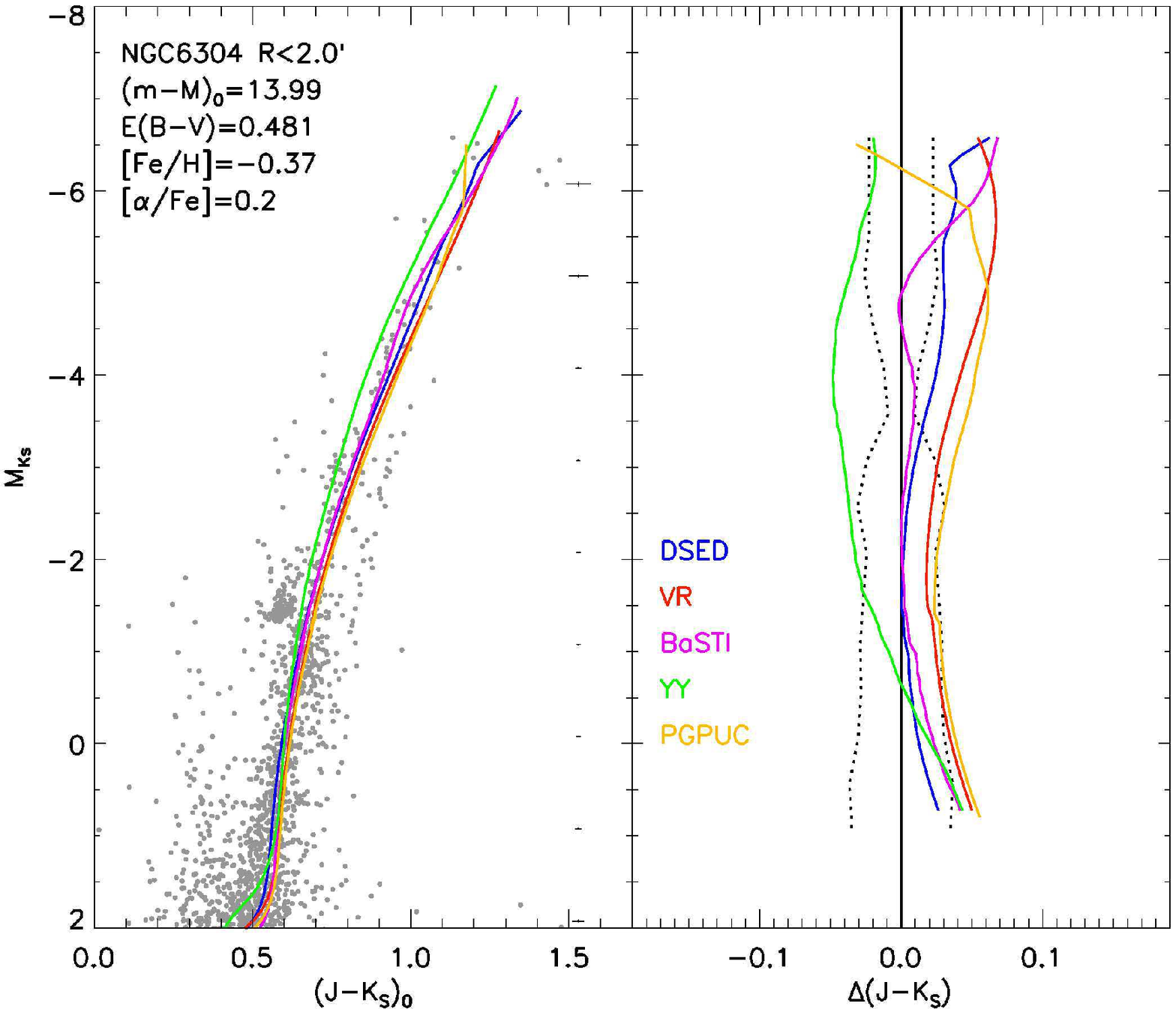}
\includegraphics[width=0.35\textwidth]{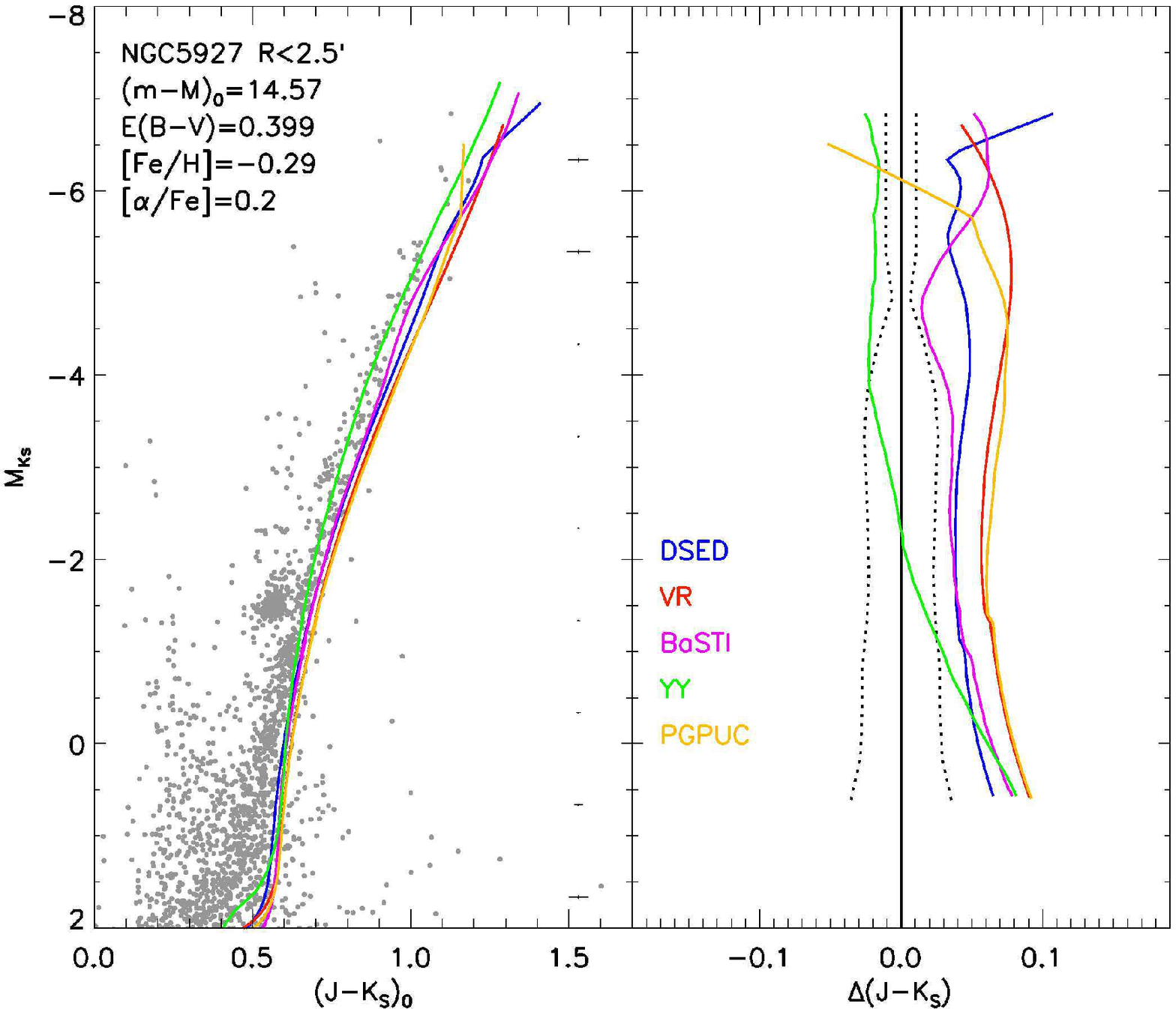}\\
\caption{Comparison between isochrones from five sets of evolutionary models 
and observed cluster photometry.  For each cluster, the left panel illustrates
the isochrones overplotted on the photometry, with median photometric errors
given along the right hand side.  The right panel illustrates the color
difference between the models and the observed fiducial sequence as a function
of absolute $K_{S}$ magnitude, where the vertical solid line represents
equality and the dotted lines indicate the estimated color uncertainty of the fiducial sequence.  Models
are color coded as indicated in the right hand panel for each cluster.}
\label{compisos_JK}
\end{figure*}

In Fig.~\ref{compisos_JK}, we compare the five sets of isochrones to 
observed ISPI photometry for eight of our target clusters 
with the highest quality photometry 
spanning the observed metallicity range.  
For increased clarity, we show two panels for each cluster, where the left
panel directly overplots the isochrones on the observed photometry, and in the
right panel we plot the color difference with 
respect to the observed fiducial sequence as a function of $M_{K}$.  For this
purpose,   
we conservatively approximate the uncertainty on the color of the fiducial
sequence as the standard deviation of the color difference between the
fiducial sequence and the observed RGB stars, calculated in moving bins
of width 0.5 mag in $K_{S}$ using exclusively stars redward of the fiducial sequence to avoid contamination from the HB and AGB\footnote{In cases
  where few stars are available (typically close to the RGB tip), the size of
  the magnitude bin was expanded until at least 5 stars were available to
  calculate the standard deviation}.  
The resulting color uncertainty as a function of magnitude is illustrated using dotted lines in the right-hand panel of Fig.~\ref{compisos_JK} for each cluster. 
We employ models with
$[\alpha/Fe]$=0.4 (except in the case of PGPUC as noted above), but for the
three most metal-rich clusters (NGC5927, NGC6304, NGC6496,
-0.46$\leq$$[Fe/H]\leq$0.29) we use models with $[\alpha/Fe]$=0.2.   This
choice is supported by both \citet{d10} and \citet{v13}, and is consistent 
with observed trends of $[\alpha/Fe]$ vs. $[Fe/H]$ as a function of
Galactocentric radius and scale height (\citealt{haydenafe} and references therein) and the
"disklike" enrichment scenario employed in earlier studies
\citep[e.g.][]{ferraro99,ferraro06}.

Total agreement, either between models and data or among various sets of
models, would be surprising due in part to
the various model-to-model differences mentioned in
Sect.~\ref{modelsect}.  However, it appears that the DSED and VR models reasonably 
reproduce the observed \textit{morphology} of the observed RGBs, although deviations in morphology
from the fiducial sequences are still seen (as with all models investigated here).  
In addition, these deviations appear to become more drastic at low ($[Fe/H]\lesssim$-2) metallicities.  While this result is fully consistent with the findings of 
\citet{brasseur} using optical-IR colors, 
we caution that at least in our case, the formal statistical
significance of these morphological deviations may not be high when 
photometric errors as well as calibration uncertainties 
are taken into account.  

To quantify the color difference
$\Delta(J-K_{S})$ between the observed RGB fiducial sequence and the five
models shown in Fig.~\ref{compisos_JK} 
for each cluster, we have calculated the mean and standard deviation 
(weighted using the observed errors shown as dotted lines in Fig.~\ref{compisos_JK}) of $\Delta(J-K_{S})$
in evenly spaced magnitude bins over the cluster RGBs ($M_{K}$$<$0).   
In this way, the mean quantifies the extent to which a model matches
observed \textit{colors}, averaged over the RGB,  
while standard deviation gauges the match in terms of CMD
\textit{morphology}.  These statistics are given for all target clusters in
Table \ref{compjktab}, and are summarized across all clusters using the median
and the median absolute deviation (MAD) in the last two rows.  The results
indicate that the DSED and VR models do the best job of reproducing the
observed CMD morphology, although BaSTI and PGPUC do nearly as well in this
sense, and the BaSTI models actually best agree with the observed 
fiducial colors \textit{in
  the mean}.

\begin{deluxetable*}{lcccccccccc}
\tablecolumns{11}
\tablecaption{Color Difference $\Delta(J-K_{S})$ Between Fiducial Sequences
  and Models}
\tablehead{ & \multicolumn{2}{c}{\textbf{DSED}} & \multicolumn{2}{c}{\textbf{VR}} &
  \multicolumn{2}{c}{\textbf{BaSTI}} & \multicolumn{2}{c}{\textbf{YY}} &
  \multicolumn{2}{c}{\textbf{PGPUC}} \\ Cluster & Mean & $\sigma$ & Mean & $\sigma$ & Mean & $\sigma$ & Mean & $\sigma$ & Mean & $\sigma$}
\startdata
NGC5927 & -0.049 & 0.018 & -0.071 & 0.009 & -0.036 & 0.016 & 0.014 & 0.033 & -0.053 & 0.028 \\
NGC6304 & -0.021 & 0.017 & -0.040 & 0.017 & -0.014 & 0.022 & 0.034 & 0.027 & -0.044 & 0.014 \\
NGC6496 & 0.002 & 0.016 & -0.017 & 0.019 & -0.004 & 0.026 & 0.039 & 0.044 & -0.023 & 0.023 \\
NGC0104 & -0.001 & 0.011 & -0.012 & 0.011 & 0.023 & 0.016 & 0.046 & 0.028 & -0.012 & 0.012 \\
NGC1851 & -0.023 & 0.008 & -0.031 & 0.007 & -0.003 & 0.020 & 0.031 & 0.040 & -0.035 & 0.006 \\
NGC2808 & -0.008 & 0.019 & -0.014 & 0.008 & 0.020 & 0.012 & 0.059 & 0.033 & -0.026 & 0.021 \\
NGC1261 & -0.017 & 0.017 & -0.032 & 0.016 & -0.007 & 0.019 & 0.026 & 0.035 & -0.033 & 0.020 \\
NGC0362 & -0.005 & 0.020 & -0.020 & 0.009 & 0.004 & 0.010 & 0.037 & 0.026 & -0.021 & 0.020 \\
NGC0288 & -0.012 & 0.008 & -0.033 & 0.021 & -0.008 & 0.033 & 0.016 & 0.054 & -0.032 & 0.012 \\
NGC6584 & -0.008 & 0.018 & -0.028 & 0.013 & -0.000 & 0.018 & 0.038 & 0.036 & -0.029 & 0.019 \\
NGC4833 & 0.006 & 0.007 & -0.016 & 0.008 & 0.013 & 0.013 & 0.057 & 0.037 & -0.015 & 0.009 \\
NGC7099 & -0.016 & 0.011 & -0.046 & 0.012 & -0.022 & 0.014 & 0.008 & 0.030 &  &  \\
\textbf{Median} & \textbf{-0.008} & \textbf{0.017} & \textbf{-0.028} & \textbf{0.012} & \textbf{-0.003} & \textbf{0.018} & \textbf{0.037} & \textbf{0.035} & \textbf{-0.029} & \textbf{0.019} \\
\textbf{MAD} & \textbf{0.009} & \textbf{0.004} & \textbf{0.012} & \textbf{0.003} & \textbf{0.011} & \textbf{0.005} & \textbf{0.011} & \textbf{0.005} & \textbf{0.006} & \textbf{0.005} \\
\enddata
\label{compjktab}
\end{deluxetable*}

We discuss practical ramifications of this result further in
Sect.~\ref{implisosect}, and now explore the effects of variations 
on $[\alpha/Fe]$ and helium predicted
by the models, in order to gain insight into the role of uncertainties 
in these quantities.

\subsection{Model Predictions: $[\alpha/Fe]$ Variations\label{afesect}}

\begin{figure}
\figurenum{8}
\label{isocomp_afe}
\includegraphics[width=0.49\textwidth]{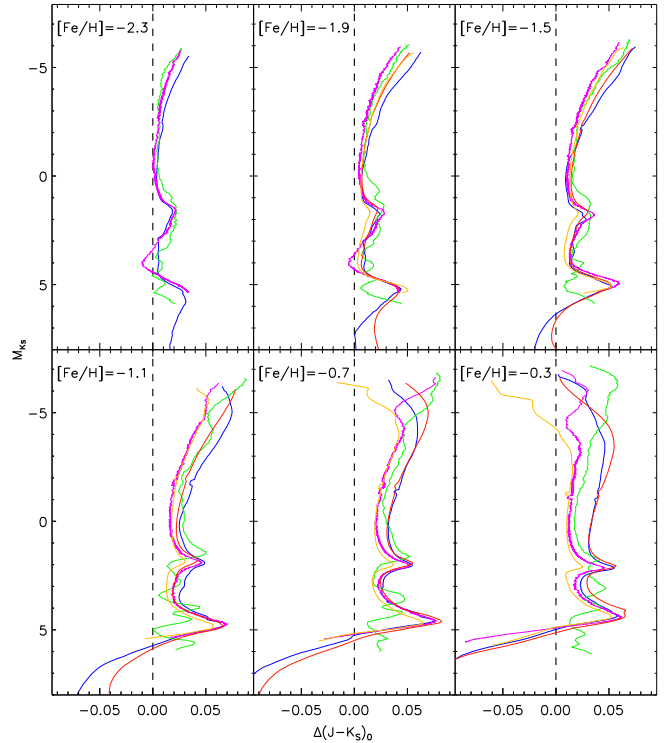}
\caption{The color difference
  between an $\alpha$-enhanced isochrone and a scaled solar isochrone for each
  model as a function of absolute $K_{S}$ magnitude.  The difference is
  plotted in the sense ($\alpha$-enhanced$-$scaled solar), and color coded by model as
  in previous figures.}
\end{figure}

The difference between scaled solar isochrones versus those which are $\alpha$-enhanced  (at fixed $Y$ and $[Fe/H]$) are illustrated in
Fig.~\ref{isocomp_afe}.  There, the $(J-K_{S})_{0}$ color difference between
isochrones with $[\alpha/Fe]$=0.4 and 0 are plotted
in the sense ($\alpha$-enhanced$-$scaled solar), for a regular grid of $[Fe/H]$ values spanning the values of the target clusters.  To illustrate the influence
of $\alpha$-enhancement in the near-IR on the main sequence, we 
extend these plots faintward beyond the lower main sequence knee (MSK).  
A couple of features are evident: First, all of the
models except for BaSTI give schematically similar predictions regarding the effect of
$\alpha$-enhancement on the upper RGB at low to intermediate
(-2$\leq$$[Fe/H]\leq$-1) metallicities.  Specifically, an enhancement of
$[\alpha/Fe]$=0.4 dex shifts the upper RGB redward by 0.05 mag or more in $(J-K_{S})$ color 
(depending on the model and the $[Fe/H]$ value), larger than seen anywhere else in the CMD with the exception of the lower ($M_{K}$$>$5) main sequence.  
Meanwhile, the
lower RGB (0$\lesssim$$M_{K}$$\lesssim$2) remains less affected save for a
slight $\sim$0.02 mag redward shift.  However, approaching solar metallicity,
the model predictions diverge somewhat, with some (DSED, VR) predicting a 
negligible effect on the RGB tip but maintaining a redward shift over the
remainder of the RGB.

\subsection{Model Predictions: Helium Enhancement\label{hesect}}

\begin{figure}
\figurenum{9}
\label{isocomp_dely}
\includegraphics[width=0.49\textwidth]{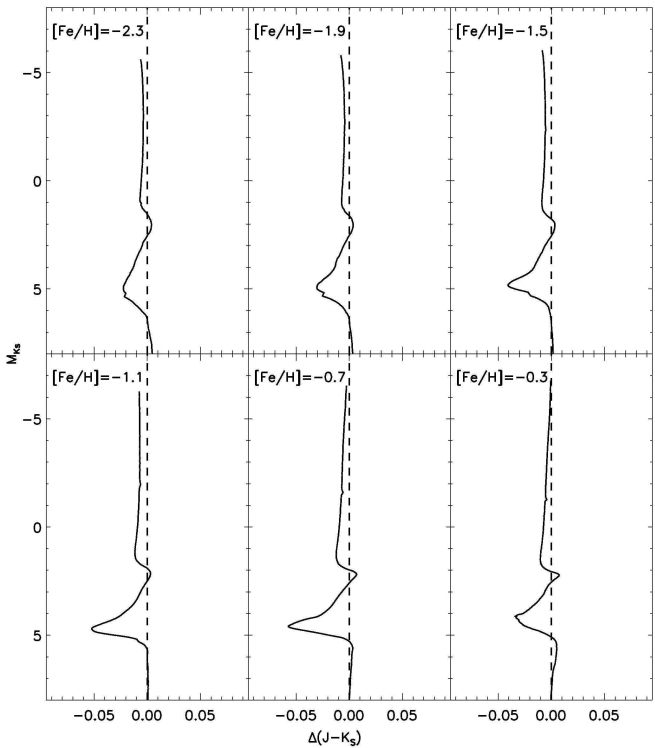}
\caption{The $J-K_{S}$ color difference
  predicted for a helium enhancement of $\Delta$$Y$=0.04 from 
  $[\alpha/Fe]$=0.4 VR models at a range of $[Fe/H]$.  The difference is
  plotted in the sense (helium enhanced-helium normal).}
\end{figure}

\begin{figure}
\centering
\figurenum{10}
\label{isocomp_cmdy}
\includegraphics[width=0.49\textwidth]{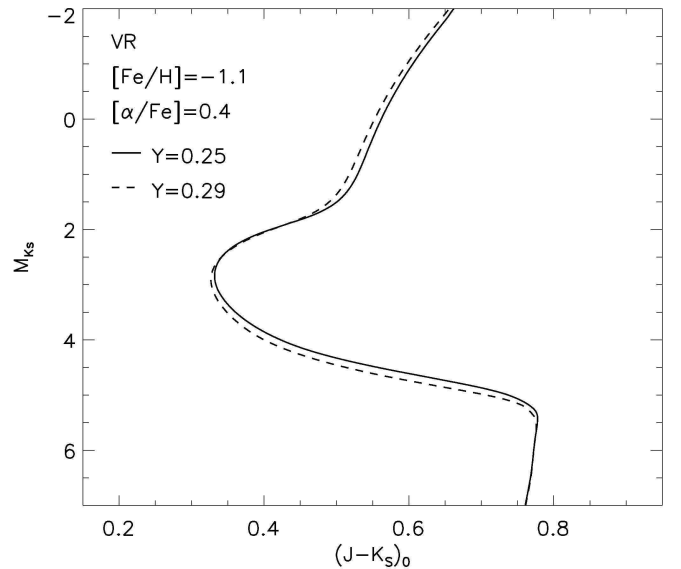}
\caption{The effect of helium enhancement in the near-IR, using VR isochrones with a fixed $[\alpha/Fe]$=0.4 and $[Fe/H]$=-1.1, shown for $Y$=0.25 (solid line) and $Y$=0.29 (dotted line).}
\end{figure}

Much effort has recently been devoted to the issue of a spread or enhancement
in helium among GGC (sub-) populations, as there are important implications for GGC formation.
Therefore we briefly examine to what extent a change in helium abundance influences
near-IR GGC CMDs.  The result of an increase of $\Delta$$Y$=0.04 
(at fixed
$[\alpha/Fe]$=0.4) is shown in Fig.~\ref{isocomp_dely}, again at a range of $[Fe/H]$ values.
Only the VR models are shown, although the results are nearly identical for other models, which is that 
near-IR colors on the RGB are minimally
affected ($\Delta$$(J-K_{S})$$<$0.01 at fixed $M_{K}$), 
with essentially no consequences for the RGB morphology.     
As it has been demonstrated that
the MSTO and MSK can be used as age indicators in the near-IR
\citep{bonoknee}, we point out that while helium enhancement shifts the MSTO
and the RGB very slightly blueward in color, 
these shifts are equivalent to $<$0.01 mag.  Similarly, the magnitude of the
MSTO and MSK are shifted faintward by nearly equal amounts ($<$0.1 mag, see
Fig.~\ref{isocomp_cmdy}).  
For this reason, a modest helium
spread or enhancement in the near-IR is of little consequence for age 
determinations employing relative CMD indices \citep[e.g.][]{v13}.
 
\subsection{Photometric Indices\label{indicessect}}

\begin{figure*}
\figurenum{11}
\plottwo{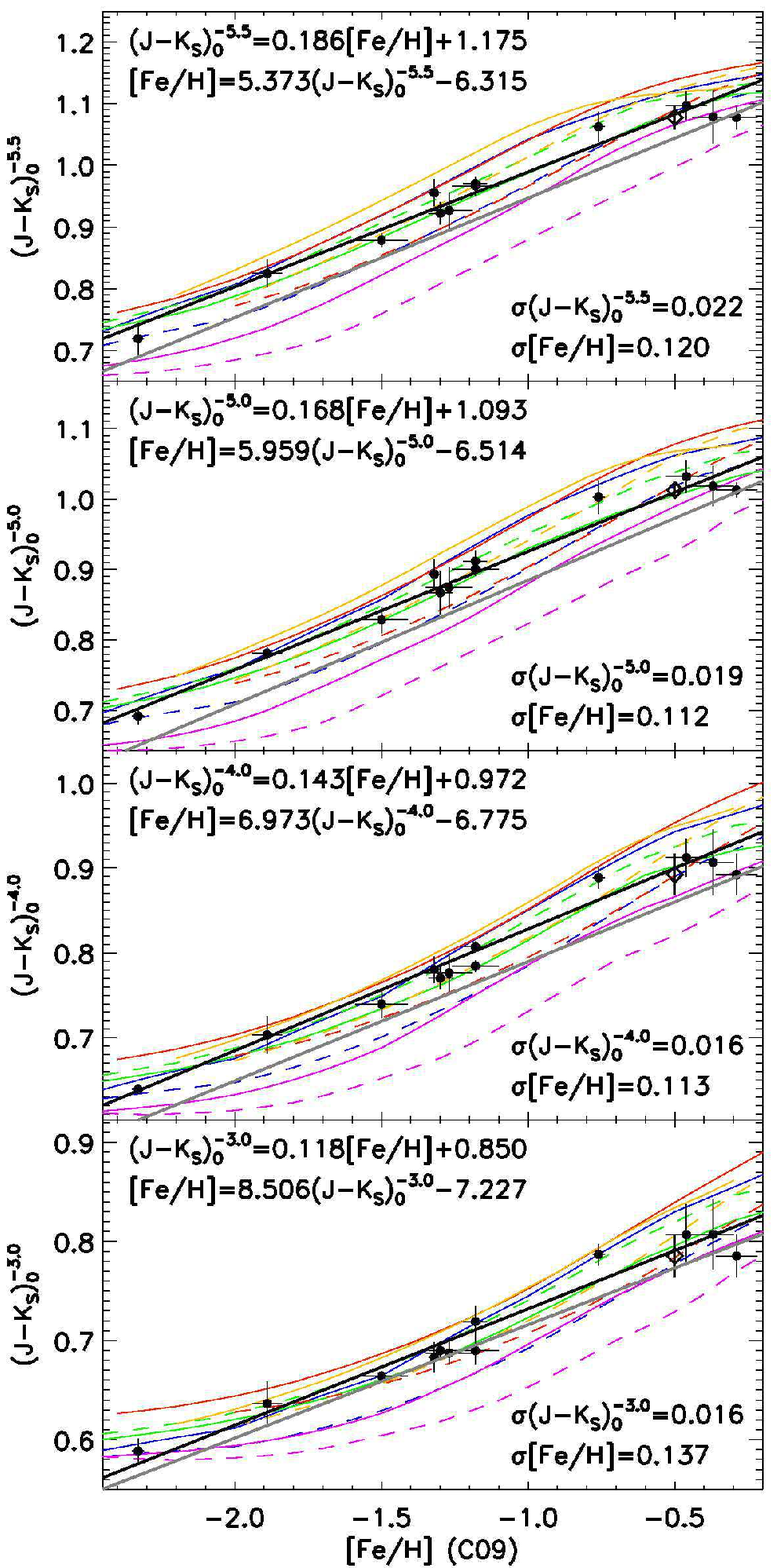}{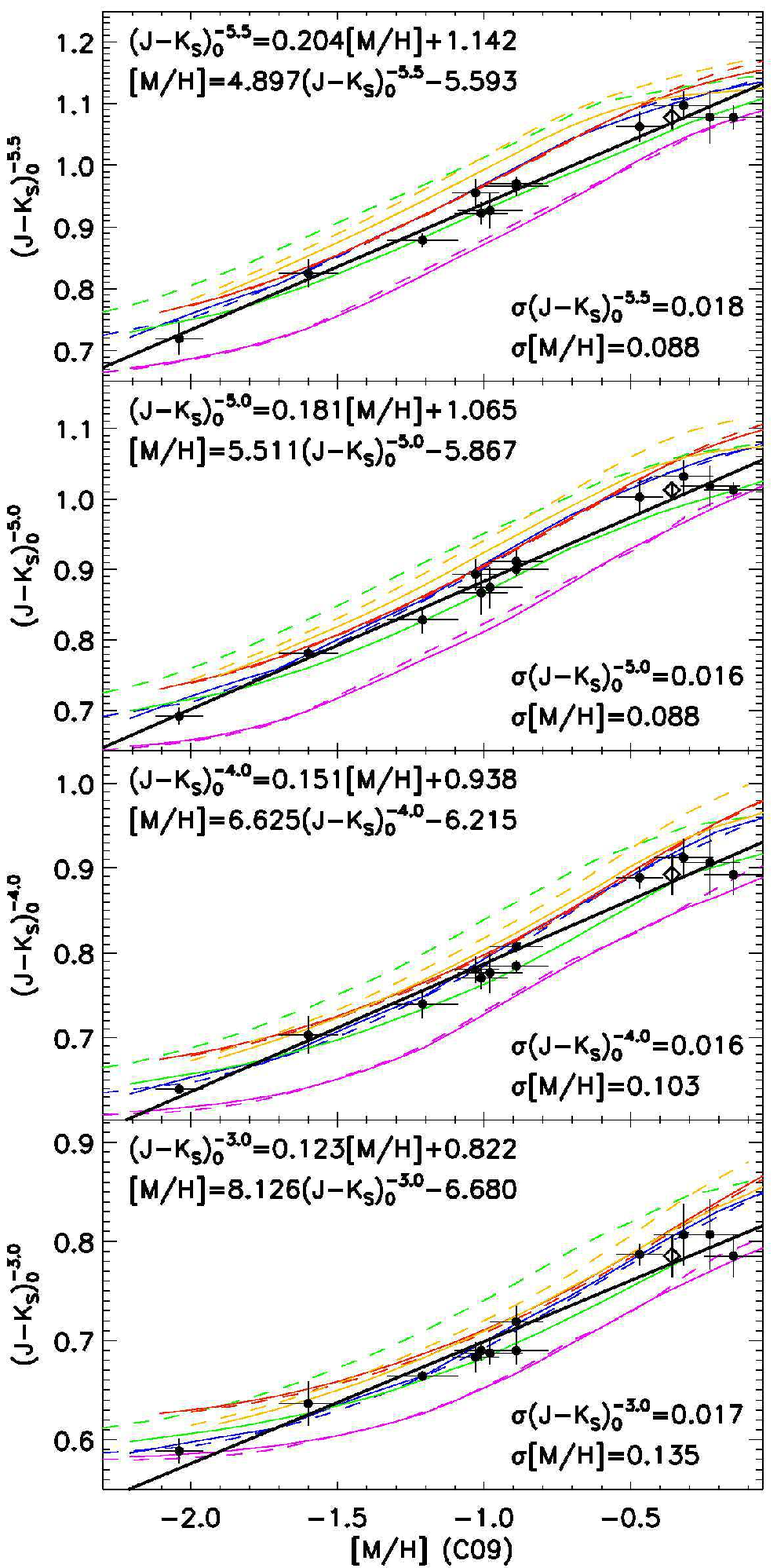}
\caption{Indices of dereddened $(J-K_{S})_{0}$ color at fixed absolute
  magnitude $M_{K}$=-5.5, -5, -4 and -3 (top to bottom) as a function of
  $[Fe/H]$ (left) and $[M/H]$ (right).  Vertical error bars
  represent measurement uncertainties only and do not include uncertainties in cluster distance and reddening (see text for details).
  Predictions of
  evolutionary models are overplotted using the same colors
  as in Fig.~\ref{compisos_JK}, where models with $[\alpha/Fe]$=0.4 are shown
 as solid lines and models with $[\alpha/Fe]$=0 are shown as dashed lines.  
  A linear fit is shown as a black solid line, with the coefficients and the
  rms deviation given in the upper left of each panel.
  The relations of \citet{v04obs}, converted to the \citet{c09}
  metallicity scale, are overplotted as grey lines in the left panels.  The open diamond represents the value for NGC 5927 assuming a metallicity of $[Fe/H]$=-0.5 (see Sect.~\ref{implisosect}).} 
\label{indicesfig}
\end{figure*}

Previous studies \citep{fcp83,cs95,ferraro00,v04obs}, cognizant of the 
fact that
the RGB becomes increasingly sensitive to metallicity at higher luminosities, 
developed a set of photometric indices to characterize the upper RGB
as a function of metallicity in the near-IR.  
These indices include (dereddened) color at fixed
(absolute) magnitude, specifically $(J-K_{S})_{0}$ at $M_{K}$=(-3,-4,-5,-5.5), as well as
absolute magnitude $M_{K}$ at $(J-K_{s})_{0}$=0.7.    
As \citet{ferraro00} point out, the use of indices constructed purely using IR
filters has the advantage of decreased sensitivity to uncertainties in 
both distance, due to the near-vertical slope of the RGB
\citep[e.g.][]{v04obs}, and reddening ($E(J-K_{S})$=0.533$E(B-V)$; \citealt{casagrande}).
These indices can be employed to measure cluster distances, reddenings and
metallicities \citep[e.g.][]{ferraro06}, and also provide another approach to
quantifying discrepancies between isochrones and observed fiducial sequences.
Therefore, in Fig.~\ref{indicesfig} we present new relations between these 
photometric indices as function of both cluster $[Fe/H]$ and global 
metallicity $[M/H]$ as
well as a comparison with the five 
sets of evolutionary models listed in Sect.~\ref{modelsect}.  
The uncertainties in dereddened color are calculated as
described in Sect.~\ref{isocompsect}, and our linear fits are performed taking
uncertainties in both axes into account\footnote{see
  \url{http://idlastro.gsfc.nasa.gov/ftp/pro/math/fitexy.pro}}.  
The resulting equations and the rms deviation of the residuals is given in
each panel of Fig.~\ref{indicesfig}, as well as inverted versions of these
equations (with $[Fe/H]$ or $[M/H]$ as the dependent variable) and the
corresponding x-axis rms.  The relations of \citet{v04obs}, transformed to the
metallicity scale of \citet{c09}, are overplotted as dotted lines for comparison.
It appears that a linear fit is a reasonable representation of $(J-K_{S})_{0}$ 
color as a function of $[Fe/H]$ and $[M/H]$, with a slope that increases
with luminosity, in general accord with previous results
\citep{ferraro00,v04obs}.

\begin{figure*}
\figurenum{12}
\plottwo{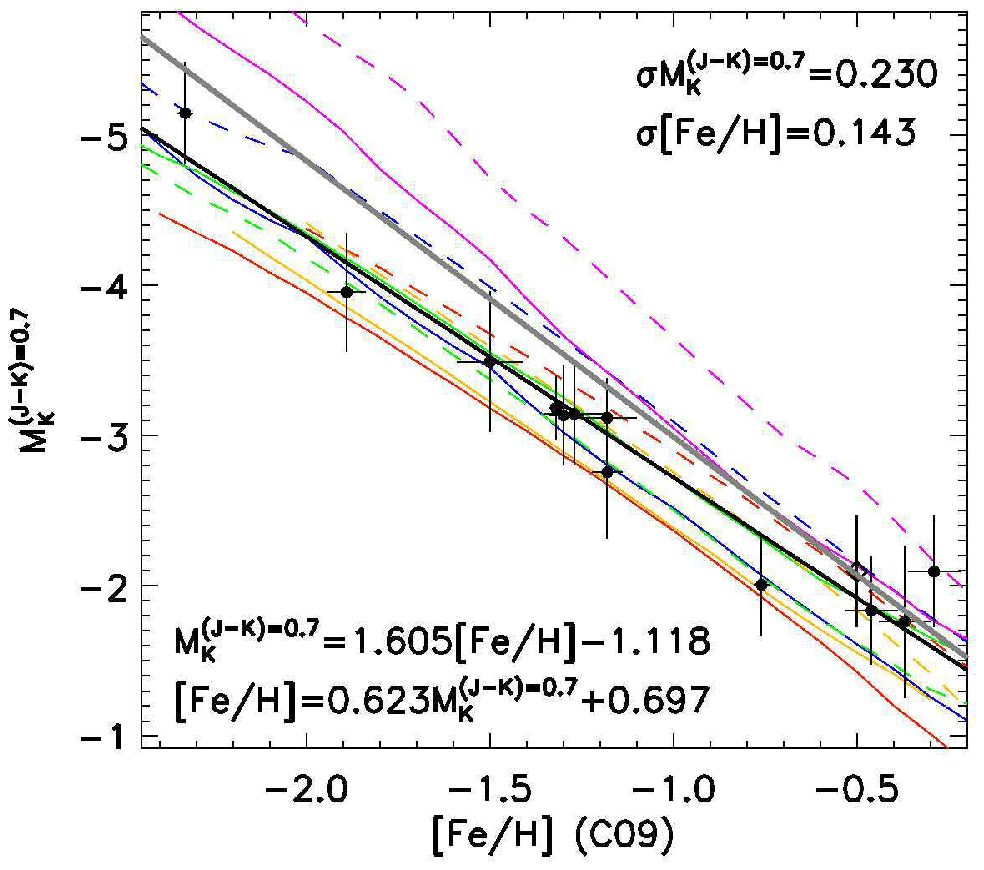}{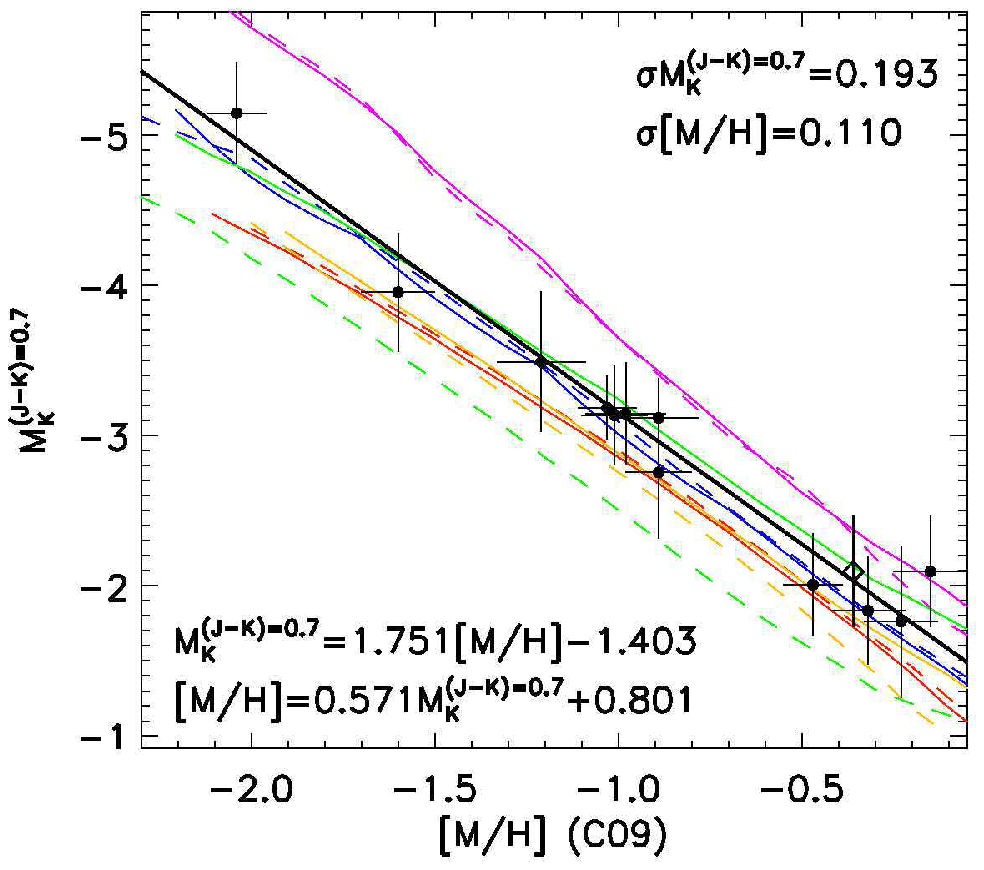}
\caption{Absolute magnitude $M_{K}$ at fixed color 
  $(J-K_{S})_{0}$=0.7 as a function of $[Fe/H]$ (left) and $[M/H]$ (right).  
  Symbols are as in Fig.~\ref{indicesfig}.}
\label{magindices}
\end{figure*}

Dereddened color at fixed absolute magnitude can also be used to characterize 
the RGB, as these indices sample a luminosity range which is dependent on
metallicity (this is apparent, for example, in Fig.~\ref{fidsabs}).  In Fig.~\ref{magindices}
we present linear fits of $M_{K}$ at fixed 
$(J-K_{S})_{0}$=0.7, 
again as a function of $[Fe/H]$ and the global metallcity $[M/H]$.  
In this case, the observational magnitude uncertainty has been calculated by
multiplying the observed color uncertainty 
by the slope of the fiducial sequence at the corresponding magnitude.  
The resulting fits are shown in Fig.~\ref{magindices}, again with
$[\alpha/Fe]$=0.4 and 0.0 models and the fits of \citet{v04obs} shown for
comparison.

Indices of color at fixed magnitude, unlike magnitude at fixed color, are
extremely sensitive to cluster reddening values, as even the most metal-rich
GGCs have RGB slopes of $|\delta(J-K_{S})/\delta$$K_{S}|$$<$0.12
\citep{v04obs,v10,chun}.  
This is an advantage when using color at fixed magnitude to describe
the RGB, since a moderate uncertainty in $[Fe/H]$ or $[M/H]$ does not
have drastic consequences for the determination of distance and reddening. 
For example, the slope of the relation in the upper left panel 
of Fig.~\ref{indicesfig} implies that $\sigma[M/H]$=0.3 dex translates to $\sigma$$A_{K}\sim$0.03. 
On the other hand, an uncertainty of
$\sigma$$E(J-K_{S})\sim$0.02, or $\sigma$$E(B-V)\sim$0.04, could account for a
scatter of $\sim$0.2 mag in the relations presented in
Fig.~\ref{magindices}, although in the present case observational
uncertainties appear to dominate.   
Interestingly, the quality of the linear fits in
Figs.~\ref{indicesfig} and \ref{magindices} improve at higher luminosities when the global
metallicity $[M/H]$ rather than $[Fe/H]$ is employed, demonstrating that
the upper RGB in the near-IR is quantifiably affected by cluster $[\alpha/Fe]$ 
consistent with the model predictions in Sect.~\ref{afesect}.  Our linear fits generally compare well to those of \citet{v04obs} given their uncertainties, and we find nearly identical slopes in the case of color at fixed magnitude (Fig.~\ref{indicesfig}).  While  zero point offsets are seen, these offsets are not large compared to the residuals of the \citet{v04obs} fits, and can likely be explained by current uncertainties in cluster distances and reddenings (discussed later in Sect.~\ref{distscalesect}).

\section{Horizontal Branch and Red Giant Branch Bump Magnitudes}

\subsection{Near-IR Horizontal Branch Magnitude\label{hbsect}}

In order to measure the magnitudes of our target cluster HBs, 
we first divide our sample into "red HB" (RHB) and "blue HB" (BHB) clusters.
This division is made
using the $\Delta$$(V-I)$ index of \citet{d10} (given in Table \ref{clusparamtab}), which 
is less vulnerable to saturation at its extremes than $(B-R)/(B+V+R)$ (e.g.~fig. 2 of \citealt{d10}).   We consider the seven clusters with 
$\delta$$(V-I)$$<$0.3 as RHB clusters, and the remainder as BHB clusters,
for the straightforward reason that all of
the RHB clusters have HB LFs which show reliably detected peaks in magnitude 
whereas the BHB clusters have HBs which do not create a
significant peak in their luminosity functions.  The HB morphology of the BHB
clusters is therefore not easily
characterized using magnitude alone due to the diagonality of BHBs in near-IR
CMDs  
(for this reason, we have included NGC 6584 among the BHB clusters despite
its value of $\delta$$(V-I)$=0.408)
so we present HB magnitudes for only the RHB clusters.

\begin{figure}
\centering
\figurenum{13}
\includegraphics[width=0.5\textwidth]{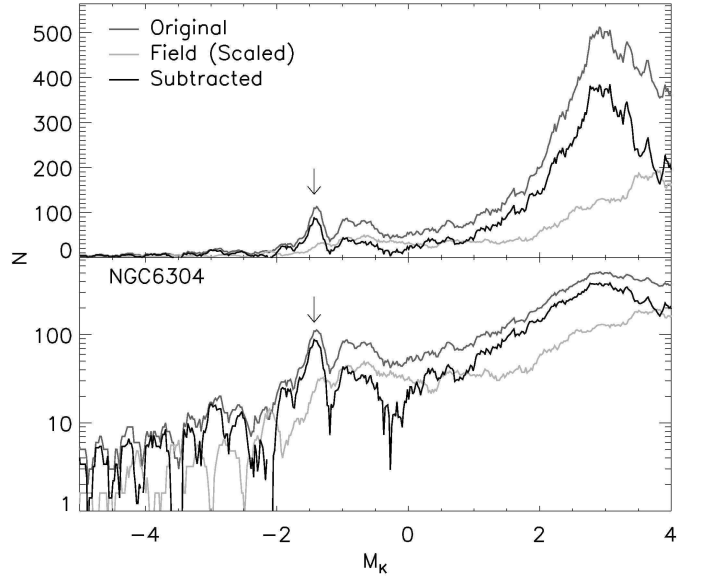}
\caption{An example luminosity function for the case of NGC
  6304, the most heavily contaminated cluster in our sample, 
  shown on a linear (top)
  and logarithmic (bottom) scale.  After constructing a multi-bin LF from the
  cluster CMD (dark grey line), 
  an LF is constructed for a comparison field, scaled by
  the ratio of cluster to field area (light grey line) and subtracted to 
  obtain the decontaminated cluster LF (black line).  The observed LF
  peak is indicated by a vertical arrow in both panels.}
\label{hblfexample}
\end{figure}

The location of the HB in $J$ and $K_{S}$ magnitude is quantified by
constructing a luminosity function (LF) from
the observed cluster CMD, 
and then identifying the location of the peak in this LF.
To mitigate the effects of binning on the final LF, 
we construct a multi-bin histogram that 
is the average of 10 individual LFs in which the bin starting points were 
shifted by 0.1 times the chosen binsize 
and these 10 LFs are averaged.  
For clusters which have at least a portion of their comparison fields outside
of their Harris (1996, 2011 revision) tidal radii (including the most heavily contaminated GGCs NGC5927, NGC6304 and NGC6496)
this procedure was repeated for the comparison field CMD, 
before scaling the field LF by the relative cluster to field area and 
subtracting it from the cluster LF.  An example is illustrated in
Fig.~\ref{hblfexample}.  
The uncertainty in the location of the LF peak is quantified using a 
simple Monte Carlo procedure wherein 
for each of 1000 iterations, the magnitude of each star in each filter is
offset by a random amount drawn from a Gaussian distribution with a 
standard deviation equal to the photometric error.  
The construction of the LFs is then repeated, 
including the generation of the multi-bin
histograms, and the location of the peak magnitude in each iteration reported.  
The uncertainty in the magnitude of the LF peak is then the standard deviation
of the reported LF peak locations over the 1000 
Monte Carlo iterations.  
Typically, this uncertainty was smaller than a resolution element of the LF
(0.01-0.03 mag), so to be conservative, 
the two values were added in quadrature to obtain the final HB magnitude 
uncertainty, given in Table \ref{hbbumptab}.

\begin{figure}
\centering
\figurenum{14}
\includegraphics[width=0.5\textwidth]{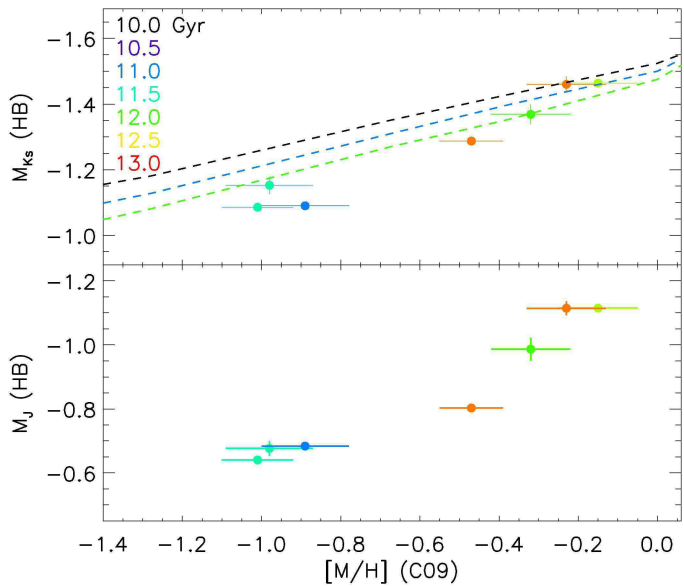}
\caption{Horizontal branch absolute magnitude in the $K_{S}$ (top) and $J$
  (bottom) filters versus cluster $[M/H]$ for RHB clusters.  In
the upper panel, the predictions of \citet{sg02} are overplotted as dashed
lines after transformation to the 2MASS photometric system \citep{carpenter}, 
and cluster and model
predictions are both color coded by age as indicated in the top left of the
upper panel.}
\label{hbmags}
\end{figure}

We plot the reported HB magnitudes as a function of $[M/H]$ in Fig.~\ref{hbmags}, color coded by 
the cluster age from \citet{d10}.  
In addition, we overplot the predictions of \citet{sg02} as dashed lines
after converting their $M_{K}$ values to the 2MASS system \citep{carpenter}, 
since this remains the only study 
directly predicting HB magnitudes in the near-IR as a function of 
cluster age and metallicity which extends to the (low metallicity, high age)
parameter space occupied by GGCs.  At the metallicity of 47 Tuc 
and higher, our results are in agreement with \cite{sg02} given
uncertainties on absolute age ($\geq$0.5 Gyr from measurement errors alone; \citealt{d10}), but
at lower metallicities, the models overestimate the $K$-band 
luminosity of the HB by as much as 0.1 mag in the case of NGC 1851.  
This issue should clearly be revisited with a larger sample of GGCs, and 
inter- and intra-cluster abundance variations not included in the \citet{sg02}
models could possibly account for this
discrepancy, but perhaps a more likely possibility is simply the remaining
uncertainty in the GGC distance and age scales (this is discussed further in
Sect.~\ref{distscalesect}).  
However, a dependence of $M_{K}(HB)$
on metallicity for GGCs has been observed at least since the study of
\citet{gs02}, and the observed HB magnitudes are not the culprit: The median 
HB magnitudes of \citet{gs02}, measured directly from 2MASS,
differ from ours by $<$0.04 mag (after accounting for the offset of
0.044 mag between photometric systems; \citealt{carpenter})
for both GGCs included in their study (47 Tuc and NGC 362).

We note that our comparison to \citet{gs02} and \citet{sg02} is not strictly 
homogenous, as
the former use the median of a CMD-selected region to characterize the HB
magnitude and the latter
give a mean HB magnitude.   
However, we found
that employing any of these alternate techniques yielded HB magnitudes
which differed from the LF peak by $\lesssim$0.02 mag in the mean, so that 
in this case, the choice of methodology
for quantifying the HB peak does not significantly impact 
our conclusions, 
although radial population gradients which vary
with HB color (for example due to light element abundance variations
cf.~\citealt{carrettanao}) could play a role.  

\subsection{The Red Giant Branch Bump (RGBB)}

\begin{figure}
\centering
\figurenum{15}
\includegraphics[width=0.5\textwidth]{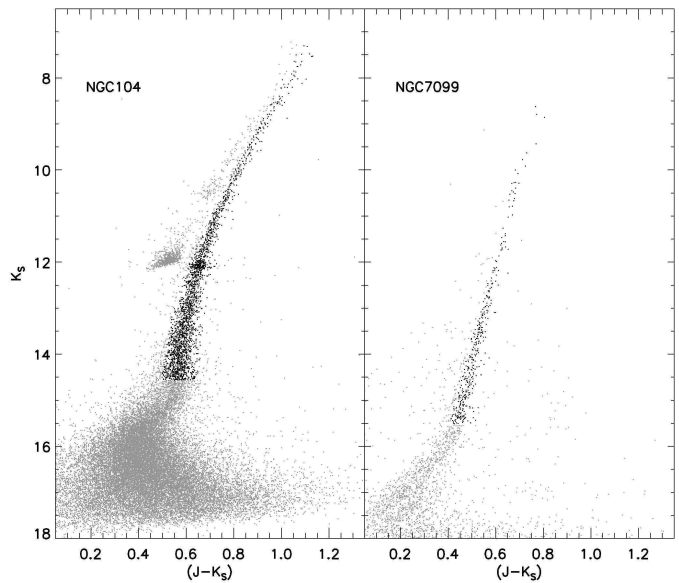}
\caption{Examples of the selection of RGB stars for the construction of the
  RGB LF, shown for the cases of NGC 104 (47 Tuc; left) and NGC 7099 (M30;
  right).  All cluster sources are shown in grey, and stars used to construct
  RGB LFs are shown in black for each cluster.}
\label{bumpcmdexample}
\end{figure}

\begin{figure}
\centering
\figurenum{16}
\includegraphics[width=0.5\textwidth]{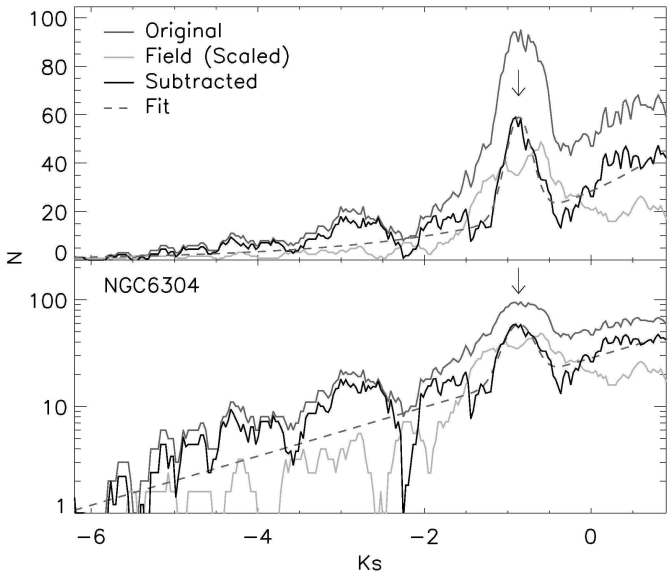}
\caption{Example $K_{S}$-band RGB LF for NGC 6304, shown on a linear (top) and
  logarithmic (bottom) scale.  Symbols are as in
  Fig.~\ref{hblfexample} and the exponential plus Gaussian fit used to
quantify the RGBB magnitude (indicated by the vertical arrow) is shown as a dotted line.}
\label{lfbumpexample}
\end{figure}

The magnitude of the RGBB is measured using a procedure similar to that of
\citet{natafbump}.  Specifically, the RGB LF of a cluster is built using stars
selected from the CMD 
(e.g.~ see Fig.~\ref{bumpcmdexample}), and multibin
histograms are used to construct the LF as described in the previous section.
Again, the field LF (constructed from an identical CMD region as the cluster LF) is subtracted
where feasible.  Next, an exponential plus Gaussian function is fit to the
resulting cluster RGB LF \citep[e.g.][]{natafbump} to
measure the location of the RGBB, and an example of this procedure is shown in
Fig.~\ref{lfbumpexample}.  
Similar to Sect.~\ref{hbsect}, Monte Carlo simulations (over 1000 iterations)
are used to quantify the uncertainty on
the RGBB magnitude in each bandpass, wherein for each iteration, each star is
offset by a Gaussian deviate of its photometric error, the multibin LF is
reconstructed, an exponential plus Gaussian function is refit, and the
resulting best fit location of the Gaussian peak is reported.  The errors reported here
were then calculated as the
quadrature sum of the fit uncertainty to the observed RGBB plus the standard
deviation of the best fit RGBB magnitude across the Monte Carlo iterations.  

\begin{figure*}
\figurenum{17}
\plottwo{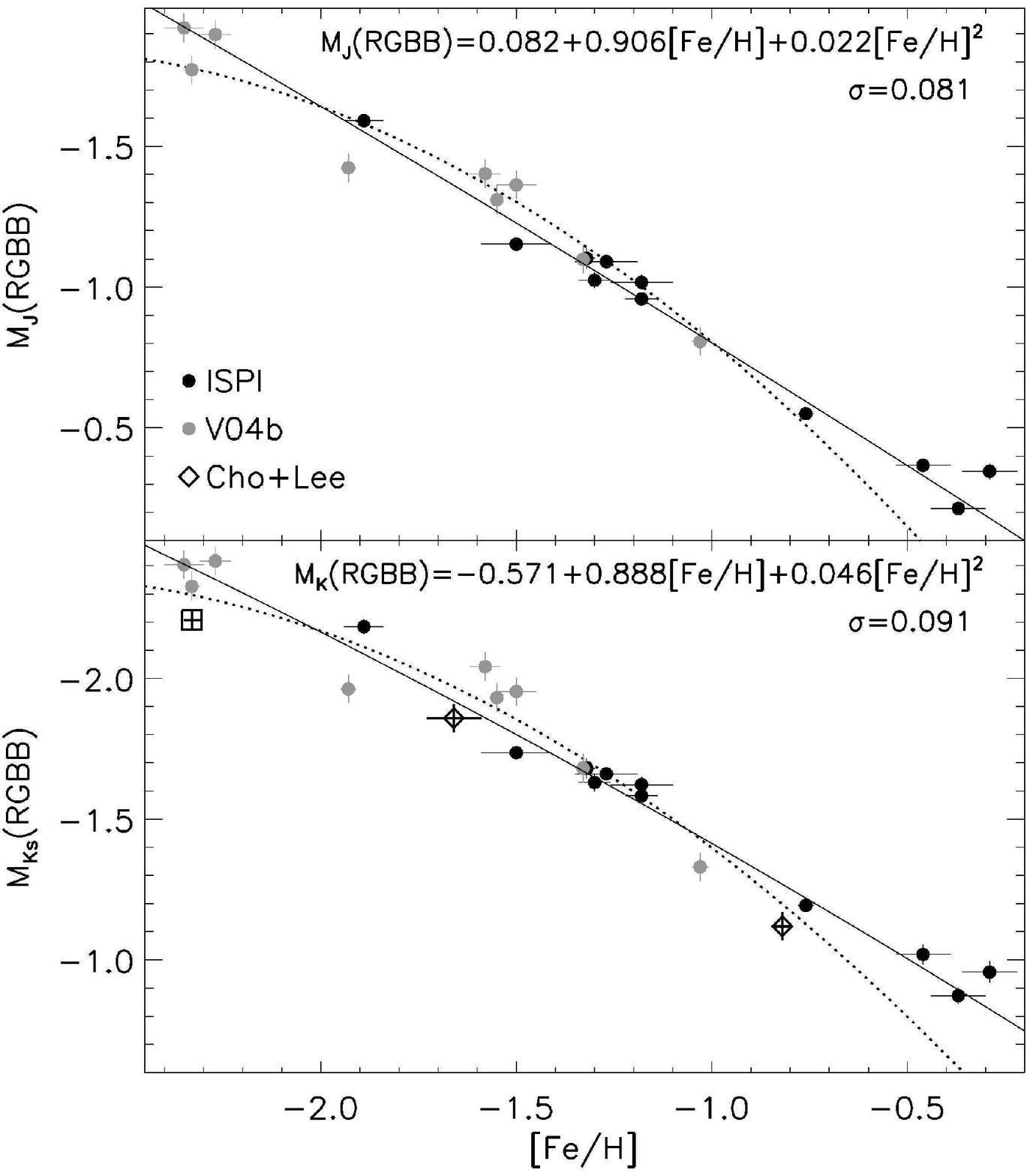}{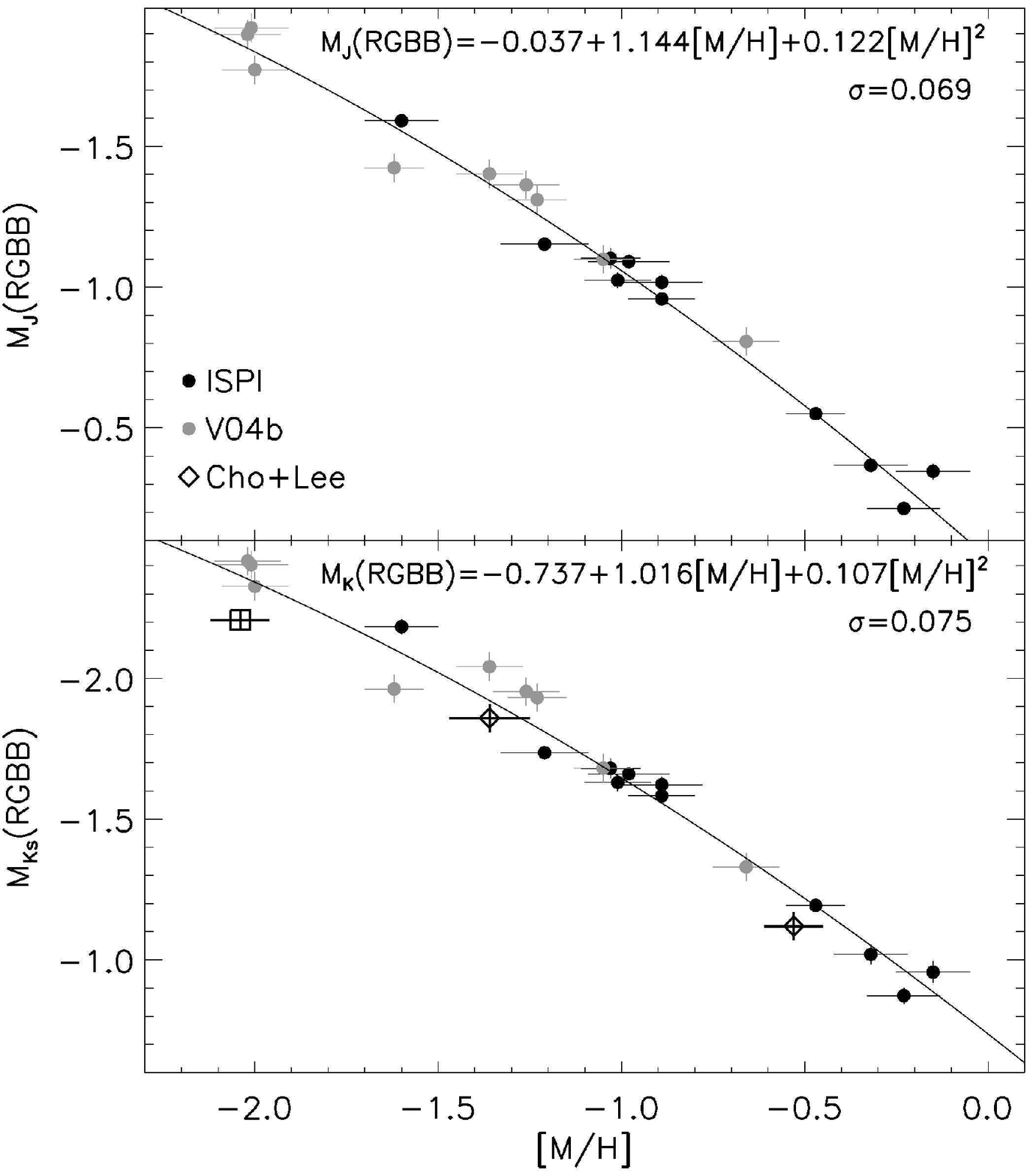}
\caption{Red giant branch bump absolute magnitude in the $J$ (upper panels)
  and $K_{S}$ (lower panels) filters versus cluster metallicity, in terms of $[Fe/H]$ (left)
  and $[M/H]$ (right).  Filled black circles represent values for ISPI target
  clusters, grey circles are from \citet{v04abs}, and black open diamonds are
  from \citet{chobump}.  The black square represents the RGBB location of NGC7099
  predicted from optical-infrared photometry (see text) and has not been included
  in the fits.  The solid line is a quadratic fit (with the
  resulting coefficients given in the upper right of each panel), and the
  dotted line represents the relation of \citet{v04abs}, transformed to the
  \citet{c09} $[Fe/H]$ scale.}
\label{bumpfig}
\end{figure*}

The resulting bump magnitudes and their uncertainties are listed in Table \ref{hbbumptab}, 
and the RGBB absolute magnitude in each filter, $M_{J}(RGBB)$ and
$M_{K}(RGBB)$ are shown in Fig.~\ref{bumpfig} as a function of both $[Fe/H]$
and $[M/H]$.  
Similarly to \citet{v04abs}, we were unable to detect the RGBB in NGC 7099.
However, by matching our near-IR data to the optical photometry by \citet{sarajedini07}
we were able to combine the $V$-band magnitude of the RGBB from \citet{natafbump}
with its $(V-K_{S})$ color to calculate the $K_{S}$ magnitude of the RGBB, shown in 
Fig.~\ref{bumpfig}. 
Next, we combined our sample with clusters
from \citet{chobump} and \citet{v04abs} which have high-quality distances,  
reddenings and spectroscopic metallicities \citep{c09,c10,d10}, 
and performed a quadratic fit for direct comparison with previous results. 
The relation of
\citet{v04abs}\footnote{The
  coefficients given by \citet{v04abs} for the $M_{K}(RGBB)-[Fe/H]$ relation
differ slightly between their fig.~2 and their eq.~3, and the former are used
in Fig.~\ref{bumpfig}.} converted to the \citet{c09} scale appears to deviate
from ours at high (near solar) metallicity, and we 
urge further multi-object spectroscopic studies of the most metal-rich GGCs to confirm this result.
However, transforming the \citet{v04abs} relation on the metallicity scale of 
\citet{cg97} to
the scale of \citet{c09} comes with the caveat that the
transformation between these scales is in fact poorly defined at high 
metallicities, as none of the 13
clusters employed to derive the linear transformation between scales 
are more metal rich than 47 Tuc (see fig.~8 of \citealt{c09}).
 
\begin{deluxetable*}{lcccc}
\tablecolumns{5}
\tablecaption{Red Giant Branch Bump and Horizontal Branch Magnitudes}
\tablehead{Cluster & \multicolumn{2}{c}{$M(RGBB)$} & \multicolumn{2}{c}{$M(HB)$} \\
  & $J$ & $K_{S}$ & $J$ & $K_{S}$} 
\startdata
NGC0104 & 12.728$\pm$0.010 & 12.072$\pm$0.009 & 12.492$\pm$0.013 & 11.979$\pm$0.013 \\
NGC0288 & 13.736$\pm$0.035 & 13.150$\pm$0.034 & & \\
NGC0362 & 13.754$\pm$0.026 & 13.136$\pm$0.029 & 14.155$\pm$0.011 & 13.680$\pm$0.011 \\
NGC1261 & 14.997$\pm$0.009 & 14.420$\pm$0.009 & 15.418$\pm$0.022 & 14.928$\pm$0.026 \\
NGC1851 & 14.424$\pm$0.025 & 13.808$\pm$0.028 & 14.775$\pm$0.012 & 14.340$\pm$0.011 \\
NGC2808 & 14.253$\pm$0.013 & 13.530$\pm$0.012 & & \\
NGC4833 & 12.917$\pm$0.014 & 12.133$\pm$0.028 & & \\
NGC5927 & 14.585$\pm$0.027 & 13.760$\pm$0.038 & 14.036$\pm$0.013 & 13.254$\pm$0.011 \\
NGC6304 & 14.211$\pm$0.017 & 13.295$\pm$0.029 & 13.577$\pm$0.022 & 12.709$\pm$0.024 \\
NGC6496 & 14.783$\pm$0.024 & 14.015$\pm$0.035 & 14.288$\pm$0.036 & 13.666$\pm$0.030 \\
NGC6584 & 14.624$\pm$0.011 & 13.999$\pm$0.010 & & \\
\enddata
\label{hbbumptab}
\end{deluxetable*}

\section{Discussion}
\subsection{Uncertainties in Cluster Parameters\label{distscalesect}}

To maximize homogeneity we consistently use the latest compilation of GGC
distances and reddenings from \citet{d10}, which are
based on fits of DSED isochrones to deep optical photometry from \citet{sarajedini07}.  While a detailed discussion of the GGC distance scale
is beyond the scope of this investigation, recent studies estimate current uncertainties
at the level of $\sim$0.10-0.15 mag \citep[][and references therein]{v13,vandenberg13}, consistent with a direct comparison between 
the distances reported by \citet{d10} versus those obtained via subdwarf fitting to
the GGC main sequences \citep{cohensxphe}.

In the current context, we revisit the comparison from \citet[][see their
fig.~12]{cohen6544} between the results of
\citet{d10} and \citet{v13} with regard to GGC distances, reddenings, and
metallicities.  This comparison is especially useful since both investigations 
employed identical photometric catalogs
from \citet{sarajedini07}, but different strategies to obtain 
cluster parameters.  On the one hand, \citet{d10} optimized 
their isochrone fits
for the unevolved main sequence and the lower RGB, using values from the 
\citet{h96} catalog as initial guesses and allowing slight
variations to $[Fe/H]$, $(m-M)_{F814W}$ and $E(F606W-F814W)$ when necessary.  
On the other hand, \citet{v13} fit model ZAHBs
to the observed lower envelope of cluster HBs to measure ages, restricting these models to the \citet{c09} $[Fe/H]$ values and a two part linear $[\alpha/Fe]$-$[Fe/H]$ relation.

\begin{figure}
\figurenum{18}
\epsscale{0.7}
\includegraphics[width=0.5\textwidth]{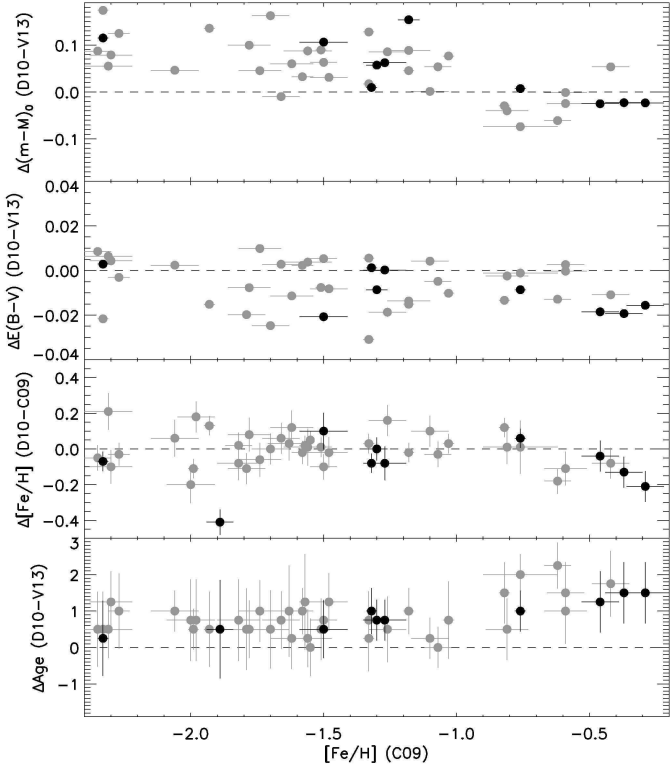}
\caption{Comparison of difference in $(m-M)_{0}$, $E(B-V)$, $[Fe/H]$ 
  and cluster age in Gyr (top to bottom respectively) between \citet{d10} and 
  \citet{v13} as a function of $[Fe/H]$
  from \citet{c09}.  The dashed line in each plot represents equality, and 
  our target clusters are overplotted in black.}
\label{compmmo}
\epsscale{1.0}
\end{figure}

In Fig.~\ref{compmmo}, we present a comparison
between the $(m-M)_{0}$, $E(B-V)$, $[Fe/H]$ and age values 
employed by \citet{d10}
and \citet{v13}.  The distance discrepancy between the two studies is
generally within the aforementioned uncertainties ($\lesssim$0.1 mag), and 
the reddening values are in particularly excellent agreement.  
We also investigate the relevance of the difference in metallicities between
the spectroscopic values reported by \citet{c09} which we employ and the 
values used by \citet{d10} for their isochrone fitting.  The two sets of
values are generally in good agreement (see Fig.~\ref{compmmo}), and
\citet{d10} discuss the role of uncertainties in $[Fe/H]$ in some detail.
However, in two cases
the metallicities employed by \citet{d10} for their isochrone fits differ
significantly from \citet{c09}:  
First, for NGC 5927 ($[Fe/H]$\citep{c09}=-0.29$\pm$0.07; $[Fe/H]$\citep{d10}=-0.5), 
the most metal rich cluster in our
sample, 
distance and reddening are 
unlikely to play a role as both $E(J-K_{S})$ and $(m-M)_{K}$ agree 
to $\leq$0.01 mag between \citet{d10} and \citet{v13}.  Furthermore,
transformations of earlier $[Fe/H]$ measurements for this cluster 
to the \citet{c09} scale 
rest fairly heavily on the value obtained by
\citet{carretta01} for NGC 6528 (see figs.~A.1 and A.2 of \citealt{c09}), 
while subsequent observations of NGC 6528
imply lower values and a substantial ($\sim$0.2 dex) spread in $[Fe/H]$ 
\citep[e.g.][]{zoccali04,sobeck06,mauro14}.   
The other significant discrepancy in cluster metallicity between \citet{c09}
and \citet{d10} is the case of NGC 4833 
($[Fe/H]$\citep{c09}=-1.89$\pm$0.05; $[Fe/H]$\citep{d10}=-2.4). 
Recent spectroscopic evidence clarifies the situation to some degree: 
On one hand, \citet{carretta4833} report
$[Fe/H]$=-2.015$\pm$0.09 dex from UVES spectra on their \citet{c09} scale, but
\citet{roederer4833} find $[Fe/H]$=-2.25$\pm$0.02 from Fe I lines and
$[Fe/H]$=-2.19$\pm$0.013 from Fe II lines, although they show that the use of
different model atmosphere grids and line analysis codes can fully account for
the discrepancy between their results and the \citet{carretta4833} value.  
Fortunately, 
the aforementioned issues 
are of little consequence for the results of 
Figs.~\ref{indicesfig}, \ref{magindices} and \ref{bumpfig}.  In fact, 
if we use values of $(m-M)_{0}$ and $E(B-V)$ from \citet{v13} and/or $[Fe/H]$
values from the \citet{d10} isochrone fits (rather than the spectroscopic
values of \citet{c09}), the fits in Figs.~\ref{indicesfig}, \ref{magindices}
and \ref{bumpfig} are unaffected to within their rms deviations. 
Alternatively, we could have simply excluded NGC 4833 and/or NGC
5927 from our fits, but this also turns out not to impact the fits beyond
their uncertainties.

Our result from Sect.~\ref{hbsect} that the \citet{sg02} models overestimate
the HB luminosity for metal-intermediate ($[M/H]\lesssim$-0.8) RHB clusters is
also robust to the
choice of distances, reddenings, and ages from \citet{v13} rather than
\citet{d10}.  This is due at least partially to the fact that although 
the absolute ages of \citet{v13} are generally younger than those of
\citet{d10}, the \textit{relative} ages of our target clusters are unchanged 
within their quoted uncertainties regardless of which set of 
\textit{absolute} ages one assumes (see Fig.~\ref{compmmo}, bottom panel).  
In fact, the use of distances, reddenings and ages from \citet{v13} rather than 
\citet{d10} actually increases the discrepancy between observed and
predicted $M_{K}$ at lower metallicities: 
Taking reported age and metallicity uncertainties into account, 
the shift required to bring observed $M_{K}(HB)$ values into accord with the
\citet{sg02} models at the 1$\sigma$ level is $\Delta$$M_{K}(HB)$=-0.15 and
-0.17 mag for NGC 362 and NGC 1851 respectively, compared to 
$\Delta$$M_{K}(HB)$=-0.05 and -0.10 mag respectively (see Fig.~\ref{hbmags}) 
using the \citet{d10} distance, reddening and age.   
However, we recall that the 
distances and reddenings given by \citet{v13} were determined
by assuming the lower envelope of the observed HB as the ZAHB (at the
high-mass end).  
This procedure comes with the risk that any dependence on metallicity 
of the magnitude difference between the model ZAHB and the \textit{observed}
lower envelope of the HB (as discussed extensively, for example, by \citealt{ferraro99} and \citealt{hbparams3}) 
could systematically bias our results.  For this
reason, the distances and reddenings of \citet{d10} appear a somewhat more
objective set of values with which to investigate a relation between HB
absolute magnitude and metallicity, in the sense that their isochrone fits, 
optimized for the cluster main sequences and lower RGB, give distances and reddenings perhaps more
``independently'' of the HB\footnote{Although maybe not in the most strict
  sense: The distance moduli in the \citet{h96} catalog which \citet{d10} use
  as initial guesses are based on an empirical $M_{V}(HB)-[Fe/H]$ relation.}.
We also caution that the interplay between HB morphology and other cluster
parameters, including age, metallicity, light element abundances, 
and helium abundance,
appears extremely complex and is under active investigation
\citep[e.g.][]{hbparams3,hbparams2,d10,hbparams1}.  
As our data are affected by central incompleteness, we cannot
exclude the possibility that radial
gradients in HB (sub-) populations \citep[e.g.][]{nataf47tuc,hbrad1,vanderbekehb}
play a role.  Therefore, a more detailed comparison between observed and
synthetic HBs is a topic better tackled using high spatial
resolution imaging of cluster cores.

\begin{figure}
\figurenum{19}
\epsscale{0.7}
\includegraphics[width=0.5\textwidth]{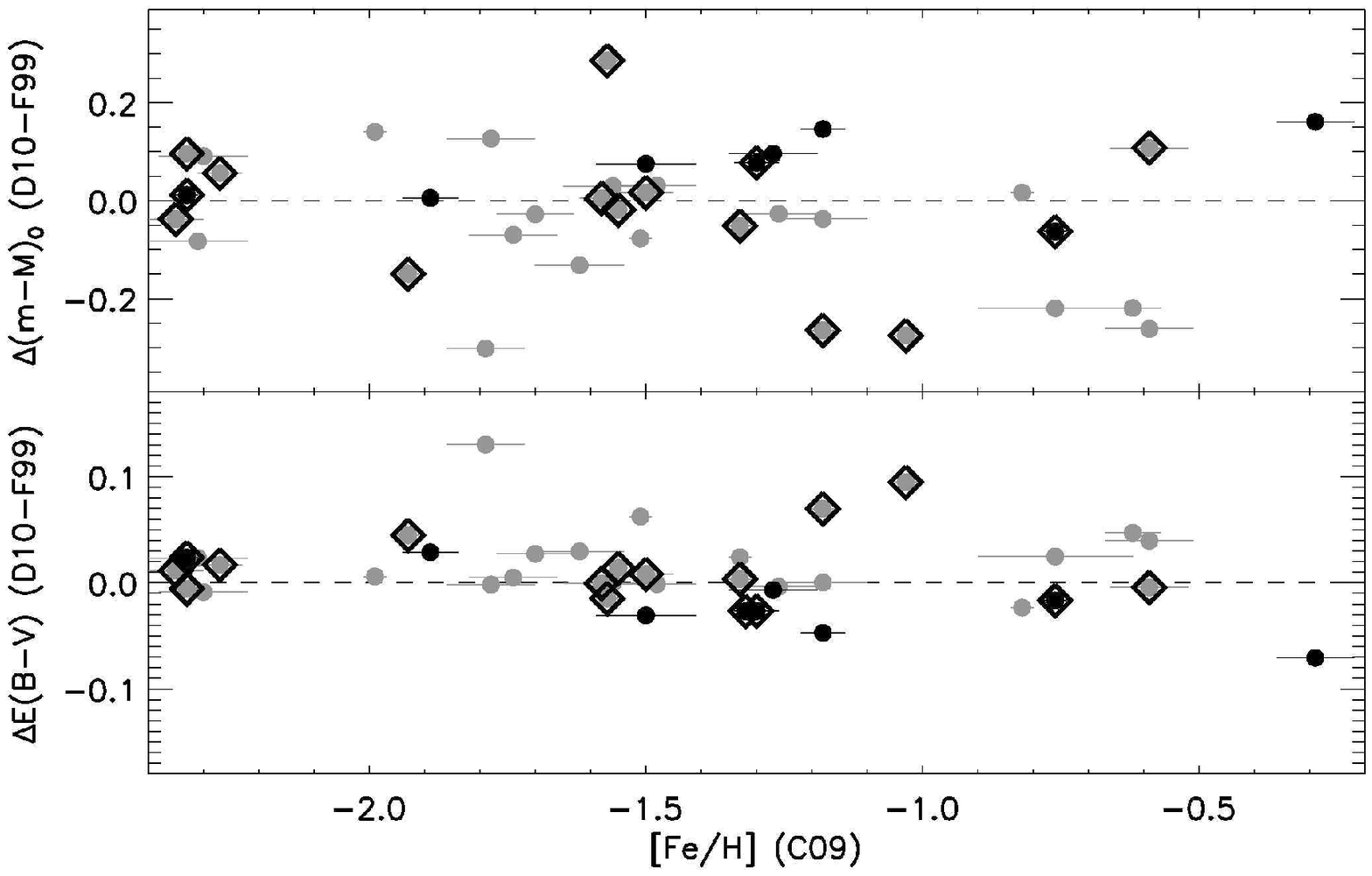}
\caption{Comparison of difference in $(m-M)_{0}$ (top) and $E(B-V)$ (bottom)
  between \citet{ferraro99} and 
  \citet{d10} as a function of $[Fe/H]$
  from \citet{c09}.  Clusters present in both \citet{ferraro99} and
  \citet{d10} are shown as grey filled circles, 
  our target clusters are overplotted as filled black circles, and calibrating
  clusters from \citet{v04obs,v04abs} with distances and reddenings given by 
  \citet{v13} are overplotted with black diamonds.  The horizontal
  dotted line represents equality.  Note the difference in
  y-axis scales as compared to Fig.~\ref{compmmo}.}
\label{compf99}
\epsscale{1.0}
\end{figure}

Differences in cluster distances and reddenings may also be responsible for 
minor differences between our fits and those of \citet{v04obs,v04abs} in
Figs.~\ref{indicesfig}, \ref{magindices} and \ref{bumpfig}. 
In Fig.~\ref{compf99}, we plot the
difference in $(m-M)_{0}$ and $E(B-V)$ between 
\citet{d10} and \citet{ferraro99},
the latter of which was employed by \citet{v04obs,v04abs} 
to calibrate their near-IR
relations.  Again, our target clusters are shown as filled black circles, and
we have overplotted the \citet{v04obs,v04abs} calibrating clusters common to
\citet{d10} with diamonds.  Unlike the comparison between \citet{d10} and
\citet{v13}, variations in cluster distance
and reddening range up to 
$>$0.1 mag in $E(B-V)$ and $>$0.2 mag in $(m-M)_{0}$ (albeit in the opposite
sense: clusters with longer distances have smaller reddening values).  The
lower reddenings of \citet{ferraro99} are a likely contributor to the  
blueward offset of the \citet{v04obs} relations in Fig.~\ref{indicesfig}
(recall that near-IR color at fixed magnitude is more sensitive to
reddening than distance), as well as the brightward offset of their relation
in Fig.~\ref{magindices}.  However, it is possible to explain 
the offset between our relations and those
of \citet{v04obs,v04abs} without invoking differences in calibrating cluster
distances and reddenings given the calibration uncertainties, measurement
uncertainties and fit quality obtained here and by \citet{v04obs,v04abs}.

\subsection{Implications for the Use of Isochrones\label{implisosect}}

\begin{figure*}
\figurenum{20}
\includegraphics[width=0.9\textwidth]{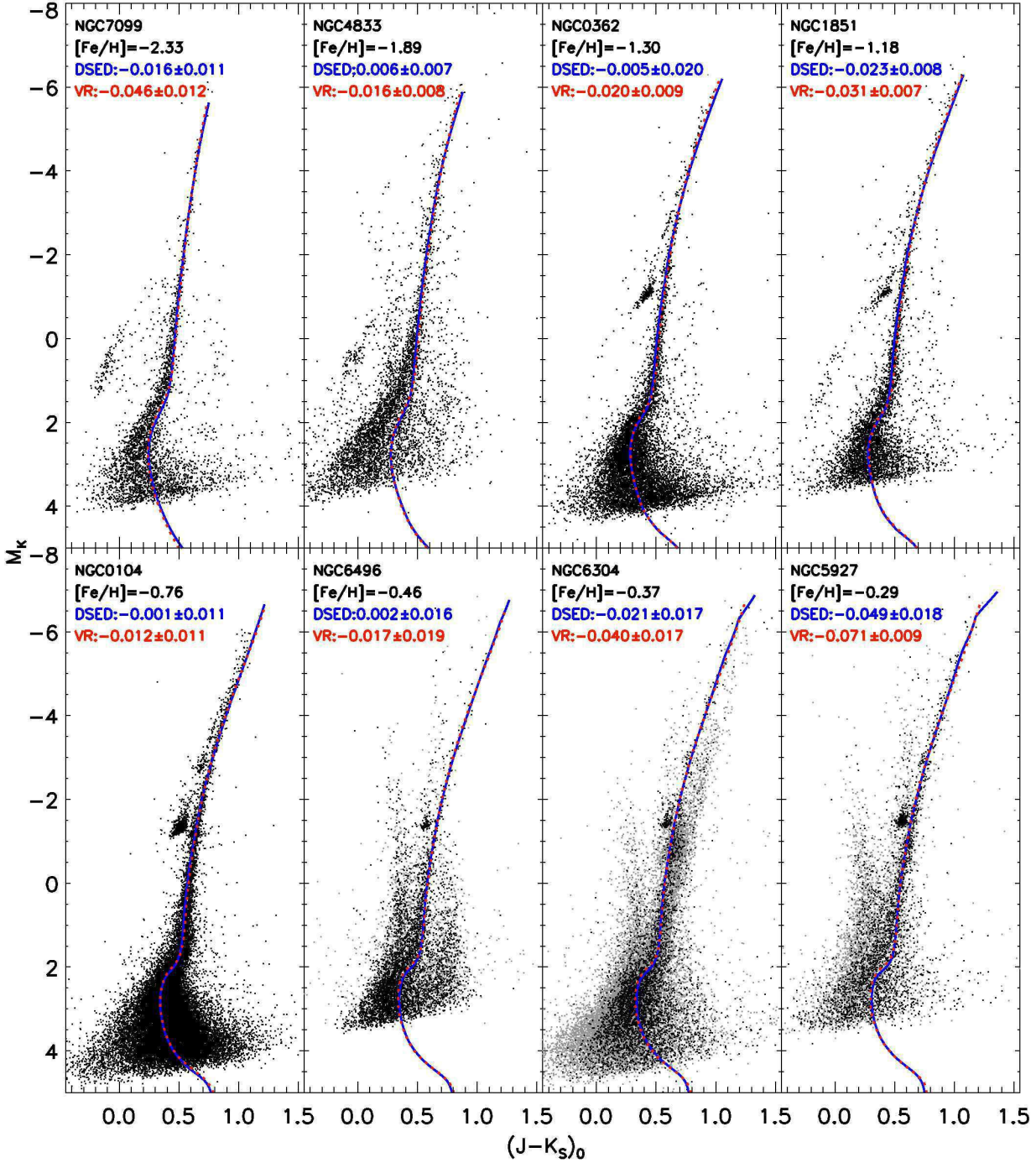}
\caption{Cluster photometry shifted to the absolute plane as in
  Fig.~\ref{compisos_JK}, with radial cuts applied in some cases as in
  Fig.~\ref{cmdsjk}.  DSED
(blue) and VR (red) models are overplotted, but after applying the fixed mean 
 color offset from Table \ref{compjktab}.  These mean offsets and their
 standard deviation are given in each 
 panel for each model.}
\label{isocomp_offset}
\end{figure*}

In Sect.~\ref{isocompsect}, we found that while models are generally offset 
redward from our fiducial sequences in $(J-K_{S})$ color, the VR and DSED
models were most successful at reproducing the \textit{morphology} of the
RGB.  In fact, if we simply apply a fixed color offset to the DSED and VR
isochrones, they reproduce the observed cluster sequences down to
the main sequence for metal-intermediate clusters, as shown in
Fig.~\ref{isocomp_offset}.  

\begin{figure}
\figurenum{21}
\includegraphics[width=0.45\textwidth]{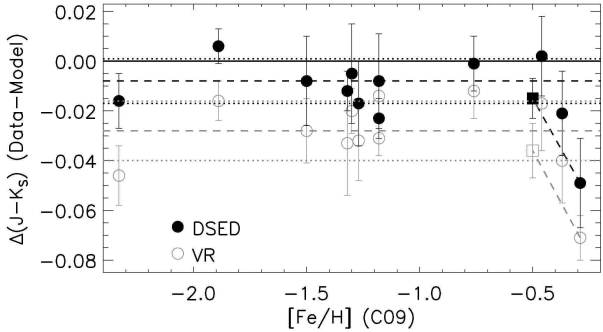}
\caption{Fixed mean color offsets applied to VR (grey open circles) and DSED
  models (black filled circles)
 in Fig.~\ref{isocomp_offset}, shown as a function of cluster $[Fe/H]$ from
  \citet{c09}.  Vertical error bars represent the standard deviation of the
  color difference between the model and observed ficucial sequence from 
  Table \ref{compjktab}.  
  Offsets for NGC 5927 calculated using $[Fe/H]$(D10)=-0.5 
  rather than $[Fe/H]$(C09)=-0.27 are shown as open and filled squares for VR and DSED
  models respectively.  The horizontal solid line represents equality, and
  horizontal dashed and dotted lines represent the median and
  median absolute deviation, respectively, 
  of the color offset across all target clusters.}
\label{offsetfig}
\end{figure}

There, we have applied the mean color offset from 
Table \ref{compjktab} to the
DSED and VR isochrones to correct for the difference between the observed 
color and that given by the isochrone.  This mean color offset and its
standard deviation (calculated as described in Sect.~\ref{isocompsect}) 
is given for each cluster in each panel of
Fig.~\ref{isocomp_offset}, and is plotted as a function of cluster $[Fe/H]$ in
Fig.~\ref{offsetfig}.  
In the case of the VR models, the blueward shifts which we find necessary
were also found at optical 
wavelengths by \citet{v13} in the sense that their ZAHB fits yielded MSTO and
RGB colors which were too red,   
but these shifts, which were all $<$0.03 mag in $(F606W-F814W)$, 
are too small alone
to explain the present color discrepancies of
$\Delta$$(J-K_{S})(VR)\sim$0.04, which translates to
$\Delta$$(F606W-F814W)(VR)\sim$0.08. 
An alternative explanation put forth by \citet{v13}
for the offset seen in optical colors is the downward revision of the GGC
metallicity scale, although this would require a shift of $[Fe/H]\gtrsim$0.3
dex based on the relations of Fig.~\ref{indicesfig}.  This is clearly unreasonable
given the quality of the spectroscopic metallicities (e.g.~Fig.~\ref{compmmo}). 

The color offsets seen in the case of the DSED models are less drastic, and
are confined to absolute values of $\Delta(J-K_{S})$ which fall well within the
margin allowed by photometric error and calibration uncertainties.  
While it may be unsurprising that the \citet{d10}
models require minimal color shifts since we adopted distances and reddenings
based on the fits of these same models at optical wavelengths, our results
serve as the first test of these models in the 2MASS filter system 
over the range of (age and metallicity) parameter space occupied by GGCs. 
For both the DSED and VR models, the largest (absolute) offset is seen for 
NGC 5927, the most metal-rich cluster in our sample.   
If we recalculate the offsets for this cluster
assuming the value of $[Fe/H]$\citep{d10}=-0.5 rather than
$[Fe/H]$\citep{c09}=-0.27   
(shown as squares in Fig.~\ref{isocomp_offset}), 
the discrepancy between models and data
is brought into the range occupied by the remainder of the GGCs, but it is not 
resolved.  Although our intention is not to sanction individual values for
cluster parameters, such a downward revision of the NGC 5927 metallicity is
suggested by the fits of 
Figs.~\ref{indicesfig}, \ref{magindices} and, to a marginal extent, Fig.~\ref{bumpfig}.
Also, 
Fig.~\ref{isocomp_offset} suggests that the DSED and VR models have
essentially identical main sequence colors at fixed metallicity, and
would continue to fall redward of the observed cluster MS in the near-IR at
the extremes of the $[Fe/H]$ range sampled, 
although deeper photometry is needed to investigate this effect
quantitatively.    
Lastly, it is possible that the color offsets between models and data could be
resolved by a shift in cluster distances.  However, throughout our
investigation we have characterized the difference between models and data in
terms of a color offset rather than a magnitude offset due to the verticality
of the RGB in the near-IR.  
For example, even a modest color offset of $\Delta$$(J-K_{S})\sim$0.02 mag 
would require a shift of $\gtrsim$0.2 mag in the GGC distance scale.

\section{Summary and Conclusions}

We have presented 2MASS-calibrated $J,K_{S}$ CMDs and fiducial sequences for
12 GGCs.  These fiducial sequences have been used to
produce relations between photometric indices which describe the shape of the
upper RGB versus cluster metallicity, in terms of both $[Fe/H]$ and $[M/H]$.  
The resulting relations have slopes in excellent agreement 
with previous studies,
and show zero point differences which can be attributed to uncertainties in the
distance and reddening of the target clusters.  While 
we have chosen to use cluster distances and reddenings based on isochrone fits
by \citet{d10} to deep optical photometry, our relations
are largely insensitive to current 
uncertainties in cluster distances, reddenings and
metallicities for the more controversial cases.  
However, their precision as well as the size of the present sample 
could be improved by detailed spectroscopic abundances for a statistically
representative quantity of cluster members, especially among 
the most metal-rich GGCs.

A comparison of empirical fiducial sequences to five sets of
evolutionary models reveals that DSED and VR models most successfully
reproduce the observed morphology of the RGB, although a color
offset is needed to reconcile models and data.  Although this offset is not
formally significant compared to 
photometric errors as well as photometric 
calibration uncertainties (at least in the case of the DSED models),
a comparison between the results of \citet{d10} and \citet{v13} suggests that it
is unlikely to be caused solely by uncertainties in 
cluster distance and reddening values.  
Models also suggest that a modest ($\Delta$$Y$=0.04)
helium enhancement negligibly affects the RGB morphology, although
$\alpha$-enhancement plays a significant role, particularly on the upper RGB.
This is in accord with our empirical relations, which generally give decreased
rms residuals when fits are performed versus the global metallicity $[M/H]$ 
rather than $[Fe/H]$.    

Relations between cluster metallicity and the near-IR
magnitudes of the RGB bump and the HB (in the case of clusters with
sufficiently red HBs) are also insensitive to the choice of recent
cluster distances, reddenings, metallicities and ages beyond current uncertainties.   
Importantly, when using the LF peak to characterize the HB magnitude and
its uncertainty, we find a
non-negligible dependence between the near-IR HB magnitude and the cluster
metallicity.  This dependence, at the level of at least 
$\delta$$M_{K}(HB)/\delta$$[M/H]$=-0.4 mag/dex depending on the choice of 
cluster parameters, is larger than that predicted by the models
of \citet{sg02} and is robust to the choice of cluster distances, reddenings
and ages.    

The photometric catalogs and observed fiducial sequences presented here 
are being made publicly available, so
that the our relations may be modified 
as the GGC distance and metallicity scales are improved.
Similarly, it is our hope that the 2MASS-calibrated photometric catalogs 
may be of use as secondary standards for near-IR adaptive optics
imagers on large telescopes, where saturation and/or a small field
of view can impede photometric calibration. 

\acknowledgements

It is a pleasure to thank Aaron Dotter for providing the results of DSED 
isochrone fits to ACS GGC Treasury Survey clusters not listed in D10, 
as well as Maurizio Salaris for sharing previous near-IR photometry of 47 Tuc 
for comparison.  
We also wish to thank Aldo Valcarce for discussions regarding updated PGPUC
isochrones, David Nataf for clarification regarding 
published $V(RGBB)$ magnitudes, and the anonymous referee for their 
insightful comments.
REC gratefully acknowledges financial support from 
Fondo GEMINI-CONICYT 32140007 and FM is thankful for financial support
from FONDECYT for project 3140177. JAG acknowledges support by the FIC-R Fund,
allocated to the project 30321072, by the Chilean Ministry of Economy through
ICM grant P071-021-F, and by Proyecto Fondecyt Postdoctoral 3130552.  
DG, MH and REC acknowledge financial support
from the Chilean BASAL Centro de Excelencia en Astrofisica y Tecnologias
Afines (CATA) grant PFB-06/2007 and MH ackowledges funding from 
ESO Comite Mixto.
This research has made use of the facilities of the Canadian Astronomy Data
Centre operated by the National Research Council of Canada with the support of
the Canadian Space Agency, as well as data products from the Two
Micron All Sky Survey, which is a joint project of the University of
Massachusetts and the Infrared Processing and Analysis Center/California
Institute of Technology, funded by the National Aeronautics and Space
Administration and the National Science Foundation.


\end{document}